\documentclass[letterpaper,twocolumn,10pt, anonymous]{article}
\usepackage{usenix-2020-09}
\usepackage{makecell}
\usepackage{appendix}
\usepackage{tikz}
\usepackage{amsmath}
\usepackage{amsmath,amssymb,amsfonts}
\usepackage{graphicx}
\usepackage{textcomp}
\usepackage{xcolor}
\usepackage{comment}
\usepackage{booktabs}
\usepackage{enumitem}
\usepackage{url}
\usepackage[normalem]{ulem}
\usepackage{algorithm}
\usepackage{algpseudocode}
\usepackage{amsmath,bm}
\usepackage{subfigure}
\usepackage{multirow}
\usepackage{pgf-pie}
\usepackage{cite}
\usepackage{etoolbox}
\usepackage{tikz}
\usepackage[sort&compress,numbers]{natbib}
\usepackage[linesnumbered,vlined, algo2e]{algorithm2e}
\usepackage{caption}
\usepackage{pgfplots}
\usetikzlibrary{intersections, positioning, calc, math, plotmarks}
\usepackage{xcolor,colortbl}
\definecolor{Gray}{gray}{0.9}

\usepackage{amssymb}
\usepackage{graphicx}
\usepackage{color}
\makeatletter
\newcommand{\btriangle}{\mathpalette\btriangle@\relax}
\newcommand{\btriangle@}[2]{%
  \begingroup
  \sbox\z@{$\m@th#1\triangle$}%
  \makebox[\wd\z@]{%
    \raisebox{0.04\height}{%
      \resizebox{1.2\wd\z@}{1.0\ht\z@}{%
        $\m@th#1\blacktriangle$%
      }%
    }%
  }%
  \endgroup
}
\makeatother
\usepackage{array}
\newcolumntype{P}[1]{>{\centering\arraybackslash}p{#1}}
\usepackage{enumitem}
\setlist[itemize]{leftmargin=*, noitemsep}
\setlist[enumerate]{leftmargin=*, noitemsep}

\usepackage{newtxmath}
\usepackage{euscript} 

\usepackage{graphicx}
\makeatletter
\DeclareFontFamily{U}{tipa}{}
\DeclareFontShape{U}{tipa}{m}{n}{<->tipa10}{}
\newcommand{\arc@char}{{\usefont{U}{tipa}{m}{n}\symbol{62}}}%

\newcommand{\arc}[1]{\mathpalette\arc@arc{#1}}

\newcommand{\arc@arc}[2]{%
  \sbox0{$\m@th#1#2$}%
  \vbox{
    \hbox{\resizebox{\wd0}{\height}{\arc@char}}
    \nointerlineskip
    \box0
  }%
}
\makeatother

\begin{document}

\title{SCVI: Bridging Social and Cyber Dimensions for Comprehensive Vulnerability Assessment}


\author{
{\rm Shutonu Mitra}\\
Virginia Tech
\and
{\rm Tomas Neguyen}\\
Virginia Tech
\and
{\rm Qi Zhang}\\
Virginia Tech
\and
{\rm Hyungmin Kim}\\
Virginia Tech
\and
{\rm Hossein Salemi}\\
George Mason University
\and
{\rm Chen-Wei Chang}\\
Virginia Tech
\and
{\rm Fengxiu Zhang}\\
George Mason University
\and
{\rm Michin Hong}\\
Indiana University
\and
{\rm Chang-Tien Lu}\\
Virginia Tech
\and
{\rm Hemant Purohit}\\
George Mason University
\and
{\rm Jin-Hee Cho}\\
Virginia Tech
}
\maketitle
%

\begin{abstract}
The rise of cyber threats on social media platforms necessitates advanced metrics to assess and mitigate social cyber vulnerabilities. This paper presents the Social Cyber Vulnerability Index (SCVI), a novel framework integrating individual-level factors (e.g., awareness, behavioral traits, psychological attributes) and attack-level characteristics (e.g., frequency, consequence, sophistication) for comprehensive socio-cyber vulnerability assessment. SCVI is validated using survey data (iPoll) and textual data (Reddit scam reports), demonstrating adaptability across modalities while revealing demographic disparities and regional vulnerabilities. Comparative analyses with the Common Vulnerability Scoring System (CVSS) and the Social Vulnerability Index (SVI) show SCVI’s superior ability to capture nuanced socio-technical risks. Monte Carlo-based weight variability analysis confirms SCVI’s robustness and highlights its utility in identifying high-risk groups. By addressing gaps in traditional metrics, SCVI offers actionable insights for policymakers and practitioners, advancing inclusive strategies to mitigate emerging threats such as AI-powered phishing and deepfake scams.
\end{abstract}

\section{Introduction}
\label{sec:intro}

\subsection{Motivation \& Goal}
The increasing digital engagement of people on social media platforms has heightened their exposure to various cyber threats. Many individuals lack the digital literacy needed to recognize and mitigate these threats, making them prime targets for cybercriminals. The consequences of scams, phishing, and other cyberattacks are severe, resulting in substantial financial loss, emotional distress, and diminished trust in online services. Although the Common Vulnerability Scoring System (CVSS) remains a widely adopted standard for assessing technical vulnerabilities in systems and networks~\cite{black2008cyber}, no prior work has introduced a dedicated index to capture social cyber vulnerabilities affecting targeted populations. Similarly, while the Social Vulnerability Index (SVI) identifies community vulnerabilities using socio-economic factors, it overlooks cyber-specific risks~\cite{flanagan2011social}. This work addresses these gaps by developing the Social Cyber Vulnerability Index (SCVI), a novel framework integrating individual-level characteristics and attack-specific threats, offering a specialized metric to inform policymakers and social media platforms about the vulnerabilities of high-risk demographics.

The importance of this work becomes evident in its effort to bridge the divide between social and technical vulnerabilities. SCVI leverages insights from studies highlighting the influence of behavioral, psychological, and sociodemographic factors on vulnerability~\cite{koning2023risk, frauenstein2020susceptibility}. Unlike traditional metrics such as CVSS or the socio-technical Cyber Risk Index~\cite{bolpagni2022cyber}, SCVI emphasizes nuanced dimensions of human susceptibility, incorporating elements like psychological traits (e.g., impulsivity, credulity), past experiences with scams, and the sophistication of attack techniques. By integrating these diverse dimensions, 
SCVI enables a comprehensive evaluation of vulnerabilities across multiple contexts.

The methodological complexity of visualizing social cyber vulnerabilities further underscores the need for SCVI. Behavioral, social, and psychological factors are inherently multi-dimensional and context-dependent~\cite{judges2017role, sur2021contextual}. Designing an index synthesizing these elements requires sophisticated analytical approaches, including sensitivity analyses and multi-dimensional visualizations. For instance, \citet{alim2011axioms} proposed metrics such as Individual Vulnerability (VI) and Relative Vulnerability (VR) for social engineering risks but lacked validation and adaptability to real-world complexities. SCVI addresses these shortcomings, offering a robust and scalable framework validated through survey (iPoll) and social media (Reddit) datasets.

Finally, this work shows that user vulnerability can be assessed using survey responses and textual data from social media. Social media interactions provide insights into manipulative attempts, deceptive offers, and exploitation patterns targeting specific demographics~\cite{prasad2023diving, edwards2018geography}. SCVI leverages these insights to highlight trends, enabling platforms to craft strategies tailored to threats like AI-powered phishing or deepfake-related scams~\cite{maddireddy2021evolutionary}. By encompassing diverse data sources, SCVI advances cybersecurity practices, ensuring inclusivity and adaptability in addressing evolving threats.

{\bf The primary goal of this work} is to develop and validate SCVI, a comprehensive framework that integrates individual-level factors (e.g., awareness, psychological traits, behavioral characteristics) and attack-level characteristics (e.g., frequency, consequence, sophistication)~\cite{nguyen2021analysis, g2024spatiotemporal}. Through diverse data sources, this study aims to enhance socio-cyber vulnerability assessment, providing actionable insights for policymakers, practitioners, and platforms to design targeted interventions and close gaps in traditional metrics.

\subsection{Key Contributions}
This work makes the following {\bf key contributions}:
\begin{enumerate}

    \item \textbf{Innovative Social Cyber Vulnerability Metric:} This research introduces the Social Cyber Vulnerability Index (SCVI). This novel framework integrates individual-level factors (e.g., awareness, behavioral traits, psychological attributes) and attack-level characteristics (e.g., frequency, consequence, sophistication). SCVI addresses gaps in traditional metrics like CVSS and SVI, offering a comprehensive tool for assessing socio-cyber vulnerabilities. Its robust methodology, including feature extraction from survey and social media data and multi-dimensional vulnerability visualization, underscores its adaptability and utility.

    \item \textbf{Validation Across Diverse Data Sources:} SCVI is validated using survey data (iPoll) and textual data (Reddit scam reports), demonstrating versatility across modalities. By incorporating individual and contextual factors, SCVI provides granular insights into cyber vulnerabilities, revealing geographical and demographic disparities that inform region-specific interventions and tailored strategies.

    \item \textbf{Rigorous Empirical Validation and Comparison:} Sensitivity and weight variability analyses validate SCVI’s reliability and robustness. Monte Carlo simulations highlight the dynamic contributions of individual and attack-level factors to SCVI variability. Comparative evaluations with CVSS and SVI confirm SCVI’s superior ability to capture nuanced vulnerabilities while aligning with CVSS for technical risks and addressing gaps in socio-contextual indices like SVI.

    \item \textbf{Practical Implications for Inclusive Cybersecurity:} SCVI supports targeted recommendations for policymakers, social media platforms, and cybersecurity practitioners by identifying high-risk groups and various scam typologies such as financial scams, phishing attacks, romance scams, online shopping fraud, tech support scams, etc. Its emphasis on demographic and sociocultural dimensions advances inclusivity. Future directions include expanding SCVI’s geographic and demographic coverage, optimizing weighting with data-driven approaches, and adapting to emerging threats like AI-powered phishing and deepfake fraud, ensuring its ongoing relevance.

\end{enumerate}

\section{Related Work} \label{sec:related-work}

\subsection{Cyber Vulnerability Metrics \& Indices}
\label{subsec:vul-metrics}

\citet{black2008cyber} evaluated CVSS for standardizing and assessing software vulnerabilities based on exploitability and impact on confidentiality, integrity, and availability. \citet{flanagan2011social} introduced the Social Vulnerability Index (SVI) to identify vulnerable communities using socio-economic factors but did not address cyber vulnerabilities. \citet{bolpagni2022cyber} proposed a socio-technical Cyber Risk Index, linking higher HDI with lower cyber risk, though it excludes social media risks. \citet{frauenstein2020susceptibility} explored personality traits influencing phishing susceptibility but overlooked cultural factors and perceived risk. While CVSS and SVI overlook nuances in social engineering and modern cyber vulnerabilities, particularly in social media and evolving online threats, the Cyber Risk Index fails to address these threats' rapid evolution. \citet{frauenstein2020susceptibility} highlighted personality traits but neglected broader contextual factors.

To contextualize cybersecurity metrics, \citet{bhol2023taxonomy} proposed a taxonomy categorizing metrics into vulnerabilities, protections, threats, users, and situations, leveraging Multi-Criteria Decision Making (MCDM) to enhance organizational cybersecurity. Similarly, \citet{van2021respite} reviewed socio-technical cybersecurity metrics for SMEs, introducing the SYMBALS method to prioritize and evaluate metrics, leading to a socio-technical framework tailored for SMEs. Further, \citet{alim2011axioms} proposed metrics like {\em Individual Vulnerability} (VI) and {\em Relative Vulnerability} (VR) to assess social engineering risks in online networks.  However, the frameworks in \cite{bhol2023taxonomy, van2021respite} may not reflect the latest advancements and can lack responsiveness to evolving threats. Although the VI and VR~\cite{alim2011axioms} are innovative, they require validation and may oversimplify real-world network complexities, limiting their practical applicability.

These gaps highlight the need for an inclusive vulnerability metric considering victim characteristics, demographics, and attack-specific factors. Existing frameworks fail to capture the evolving nature of social cyber threats, underscoring the importance of specialized tools to better protect vulnerable populations.

\subsection{AI-Driven Cybersecurity Solutions}

\citet{maddireddy2021evolutionary} proposed a multi-modal AI framework integrating machine learning, deep learning, and natural language processing (NLP) for ransomware detection. Evaluated on datasets like VirusShare and UNSW-NB15, it achieved high detection rates and adversarial robustness. \citet{jamil2018mpmpa} introduced the MPMPA model to mitigate phishing on Facebook, employing finite-state machines for detection and real-time alerts validated through realistic scenarios. \citet{prasad2023diving} developed {\em SnorCall}, using semi-supervised learning for analyzing illegal robocalls, achieving high labeling accuracy (90 - 100\%) and insights into scam operations. These studies demonstrated innovative AI applications for specific cybersecurity threats.  However, limitations exist. The framework in~\cite{maddireddy2021evolutionary} faces a lack of scalability and interpretability, while MPMPA~\cite{jamil2018mpmpa}’s focus on Facebook limits its generalizability. {\em SnorCall}~\cite{prasad2023diving}'s reliance on transient data and manual labeling introduced biases and ethical concerns. These limitations reflect the broader challenges in AI-driven cybersecurity, such as scalability, model interpretability, ethical considerations, and adaptability to dynamic threat landscapes, underscoring the need for further research and development to enhance robustness and usability.

\subsection{Key Predictors of Vulnerability}

\citet{koning2023risk} used Dutch survey data to examine socio-demographics, personality traits, and internet activities in fraud victimization, identifying higher risks for younger individuals, frequent internet users, and those with lower self-control, with openness to experience increasing susceptibility. \citet{sur2021contextual} analyzed older adults, showing well-being and cognitive ability reduced fraud risk, while negative life events and loneliness increased it. \citet{williams2023demographic} linked credit card fraud to financial confidence, training, marital status, and homeownership. \citet{judges2017role} found lower cognitive ability and reduced honesty-humility as key predictors among older adults. However, their works are limited to reliance on post-fraud measurements~\cite{koning2023risk}, self-reported data~\cite{sur2021contextual, williams2023demographic}, and small, non-diverse samples~\cite{judges2017role}. These findings show the role of cognitive ability, personality traits, and socio-demographic factors in understanding vulnerability to social cyber threats, underscoring the need for broader, more robust studies.

\subsection{Geographic, Demographic, and Temporal Dimensions of Social Scam Fraud}

\citet{edwards2018geography} analyzed 5,402 online dating scam profiles, revealing Nigeria as a major source (30\%) and identifying text/image reuse patterns via IP geolocation. \citet{nguyen2021analysis} examined crime patterns in India using regression analysis, finding links between demographic factors and regional disparities. \citet{g2024spatiotemporal} investigated U.S. consumer financial fraud trends (2018–2022), noting spatial clustering of fraud, especially post-COVID-19, using techniques like Moran’s I. However, they could not resolve issues with dataset biases and proxy use~\cite{edwards2018geography}, relied on secondary data lacking qualitative insights~\cite{nguyen2021analysis}, and encountered underreporting and oversimplification in spatial analyses~\cite{g2024spatiotemporal}. Despite these issues, they underscored the need for comprehensive methodologies to address fraud's complex dynamics.

The Social Cyber Vulnerability Index (SCVI) introduced here bridges these gaps by integrating data sources like social media and surveys. It offers a context-aware framework to assess vulnerabilities across diverse contexts, transcending prior limitations in adaptability and inclusivity.

\section{Measuring Social Cyber Vulnerabilities} \label{sec:proposed}

\subsection{Key Vulnerability Factors to Social Scams}
\label{subsec:key-vul-factors-sc}

Social scams represent a specialized category of online fraud~\cite{kavrestad2018defining}, targeting victims through interactive channels such as email, websites, social media, and chat rooms. While online fraud typically leverages digital mediums to facilitate fraudulent activities, social scams focus on exploiting personal relationships, social dynamics, and trust~\cite{guo2020online}. This manipulation often involves deceptive communications or persuasive tactics, ultimately aiming to defraud individuals by undermining their behavioral and psychological defenses~\cite{susser2019online}.

This work considers the following factors that shape an individual’s susceptibility to social cyberattacks. These factors align with the dimensions incorporated into the proposed {\em Social Cyber Vulnerability Index} (SCVI), which can be the basis for assessing risk to a given social scam:
\begin{enumerate}[leftmargin=*]
\item \textbf{Individual Awareness and Knowledge:} This captures how well individuals understand social cyber threats and the protective measures available. Limited awareness and knowledge of specific scam types or associated security mechanisms heighten vulnerability~\cite{albladi2020predicting}.

\item \textbf{Behavioral Patterns:}  This factor encompasses an individual's online activities and security practices. Frequent exposure to risky platforms or inadequate protective behaviors can raise the likelihood of victimization~\cite{reyns2013security}.

\item \textbf{Psychological Factors:}  This considers traits like trust and risk perception. Excessive trust in unfamiliar communications or a diminished sense of risk often translates into heightened susceptibility to scams~\cite{Robb2023}.

\item \textbf{Past Experience:} This reflects prior encounters with social cyberattacks, including the individual’s responses and recovery strategies. Experiences can either strengthen resilience or, if inadequately addressed, leave persistent vulnerabilities~\cite{fredrick2021resiliency}.

\item \textbf{Frequency of Attacks:} This refers to the number of scam attempts a user encounters. A higher frequency increases the likelihood of at least one attack succeeding. Moreover, repeated scam exposure can desensitize individuals, potentially reducing their vigilance over time~\cite{whitty2020there}.

\item \textbf{Consequences of Attacks:} This involves significant financial or emotional harm that can exacerbate overall vulnerability, as high-stakes repercussions often leave lasting negative impacts~\cite{norris2019psychology}.

\item \textbf{Sophistication of Attacks:} This assesses how convincingly a scam mimics legitimate communications or exploits personal information. Highly sophisticated (or realistic) social cyberattacks are more difficult to detect, elevating the victim's susceptibility~\cite{collier2022sophisticated}.
\end{enumerate}

By examining these interrelated factors, the SCVI provides a holistic measure of an individual’s susceptibility to social cyberattacks.  While numerous factors can contribute to vulnerabilities in social scams, this study focuses on seven key elements identified in existing research. We assess these factors using survey and social media datasets. 

\subsection{Social Cyber Vulnerability Index (SCVI)}
We propose the {\em Social Cyber Vulnerability Index} (SCVI), assessing susceptibility to cyber social scams by evaluating individual vulnerability and attack severity factors.  

The SCVI is formulated by:
\begin{eqnarray}
\mathrm{SCVI}_{i, k} = \alpha \cdot \mathrm{IVI}_{i, k} + \beta \cdot \mathrm{ASI}_{i, k}.
\end{eqnarray}
The $\mathrm{IVI}_{i, k}$ represents the Individual Vulnerability Index, indicating the vulnerability of an individual $i$ to an attack $k$, while $\mathrm{ASI}_{i, k}$ denotes the Attack Severity Index, representing the impact of attack $k$ on individual $i$. The weights $\alpha$ and $\beta$ are assigned to IVI and ASI, respectively, with $\alpha + \beta = 1$.

$\mathrm{IVI}_{i, k}$ is given by:
\begin{eqnarray}
\mathrm{IVI}_{i, k} = w_{A_{i, k}} \cdot \mathrm{A}_{i, k} + w_{B_{i, k}} \cdot \mathrm{B}_{i, k}+ \\ w_{P_{i, k}} \cdot \mathrm{P}_{i, k} + w_{E_{i, k}} \cdot \mathrm{E}_{i, k},   \nonumber       
\end{eqnarray}
where $\mathrm{A}_{i, k} = \mathrm{A}_{i, k}^A + \mathrm{A}_{i, k}^K$,  $\mathrm{B}_{i, k} = \mathrm{B}_{i, k}^R + \mathrm{B}_{i, k}^S$, $\mathrm{P}_{i, k} = \mathrm{P}_{i, k}^C + \mathrm{P}_{i, k}^I$, $\mathrm{E}_{i, k}=\mathrm{E}_{i, k}^E + \mathrm{E}_{i, k}^R$, and $w_{A_{i, k}} + w_{B_{i, k}} + w_{P_{i, k}} + w_{E_{i, k}} = 1$.  

The lack of individual $i$'s {\bf awareness and knowledge} ($\mathrm{A}_{i, k}$) includes unfamiliarity with social cyberattack $k$ ($\mathrm{A}_{i, k}^A$) and knowledge of protective measures against $k$ ($\mathrm{A}_{i, k}^K$). {\bf Behavioral patterns} ($\mathrm{B}_{i, k}$) involve the frequency of risk-enhancing behaviors ($\mathrm{B}_{i, k}^R$) and the use of security practices ($\mathrm{B}_{i, k}^S$). {\bf Psychological factors} ($\mathrm{P}_{i, k}$) include trust in communications related to $k$ ($\mathrm{P}_{i, k}^C$) and risk perception and impulsivity ($\mathrm{P}_{i, k}^I$). {\bf Experience} ($\mathrm{E}_{i, k}$) reflects past encounters with $k$ ($\mathrm{E}_{i, k}^E$) and responses to such incidents ($\mathrm{E}_{i, k}^R$).

$\mathrm{ASI}_{i, k}$ is defined by:
\begin{gather}
\mathrm{ASI}_{i, k} = w_{F_{i, k}} \cdot \mathrm{F}_{i, k} + w_{C_{i, k}} \cdot \mathrm{C}_{i, k} + w_{S_{i, k}} \cdot \mathrm{S}_{i, k}, 
\end{gather}
where $\mathrm{F}_{i, k} = \mathrm{F}_{i, k}^{TA} + \mathrm{F}_{i, k}^{AA}$, $\mathrm{C}_{i, k} = \mathrm{C}_{i, k}^{FI} + \mathrm{C}_{i, k}^{PI} + \mathrm{C}_{i, k}^{SI}$, $\mathrm{S}_{i, k} = \mathrm{S}_{i, k}^C + \mathrm{S}_{i, k}^{SE}$, and $w_{F_{i, k}} + w_{C_{i, k}} + w_{S_{i, k}} = 1$, respectively.

The impact of the {\bf frequency of attack} $k$ ($\mathrm{F}_{i, k}$) is determined by the frequency of attempted attacks ($\mathrm{F}_{i, k}^{TA}$) and the frequency of actual attacks encountered ($\mathrm{F}_{i, k}^{AA}$). The {\bf consequence of a successful attack} $k$ on individual $i$ ($\mathrm{C}_{i, k}$) is represented by its financial impact ($\mathrm{C}_{i, k}^{FI}$), emotional or psychological impact ($\mathrm{C}_{i, k}^{PI}$), and impact on personal safety ($\mathrm{C}_{i, k}^{SI}$). The {\bf sophistication of attack} $k$ ($\mathrm{S}_{i, k}$) as perceived by individual $i$ consists of the perceived degree to which the attack mimics legitimate communications ($\mathrm{S}_{i, k}^C$) and the use of personalized information or advanced social engineering techniques ($\mathrm{S}_{i, k}^{SE}$).

We consider the scale for each component (e.g., $\mathrm{A}_{i, k}$) to be in $[0, 5]$, and its value will be represented as an integer. The $\mathrm{IVI}_{i, k}$ can be captured based on a Likert-scale (i.e., from 0 to 5) questionnaire or derived from an online user's behavioral or lexical characteristics in social media. The $\mathrm{ASI}_{i, k}$ can be represented by the key characteristics of each attack perceived by online users~\cite{zong2019analyzing}. The assigned weights for IVI and ASI balance individual vulnerability and attack severity, ensuring SCVI's effectiveness for assessment and targeted interventions by integrating key factors across both dimensions.

\section{Estimating SCVI}

This section details the estimation of SCVI using two datasets: survey data from the iPoll dataset and social media data from Reddit scam reports. Integrating iPoll’s survey-based measures of risky behaviors and protective knowledge with Reddit’s user-generated scam reports, the SCVI achieves both statistical rigor and contextual depth, resulting in a comprehensive and adaptable measure of social cyber vulnerability. Together, these two datasets form a complementary foundation for SCVI estimation.

\subsection{SCVI Using the iPoll Dataset}

To compute the Social Cyber Vulnerability Index (SCVI), we utilize the survey data from the iPoll dataset \cite{ipoll-dataset2020}. The dataset, The Impostors: Stealing Money, Damaging Lives. An AARP National Survey of Adults 18+ (iPoll) thoroughly examines scam awareness, experiences, and behaviors among 4,596 adults across the United States. Conducted by NORC at the University of Chicago between January 2 and January 16, 2020. The dataset captures diverse aspects of online behavior, personal traits, and specific scams, such as romance scams, government impostor scams, and identity theft, across 113 variables. It employs computer-assisted telephone interviews (CATI) and web-based surveys, offering robust regional and national insights. While weighting factors are provided for analysis accuracy, the dataset highlights varying levels of scam susceptibility and awareness among different demographics, making it a critical resource for understanding and mitigating cyber vulnerabilities.

\paragraph{\bf Modeling the Individual Vulnerability Index (IVI).} IVI is modeled from the iPoll dataset by encoding and aggregating participant responses across seven dimensions reflecting cyber vulnerability. These dimensions include lack of Awareness ($\mathrm{A}_{i, k}^A$),lack of Knowledge of Protective Measures ($\mathrm{A}_{i, k}^K$), Frequency of Risk-Enhancing Behaviors ($\mathrm{B}_{i, k}^R$), Trust Level ($\mathrm{P}_{i, k}^C$), Risk Perception and Impulsivity ($\mathrm{P}_{i, k}^I$), Past Encounters ($\mathrm{E}_{i, k}^E$), and Responses to Past Incidents ($\mathrm{E}_{i, k}^R$). Each dimension is calculated by mapping survey responses to numerical scores based on predefined encoding schemes. For example, in $\mathrm{A}_{i, k}^A$, a response of "Very concerned" is assigned a score of 0 (indicating high vulnerability), whereas "Not at all concerned" receives a score of 3. Similarly, responses like "False" in $\mathrm{A}_{i, k}^K$ are assigned a score of 5, reflecting higher vulnerability due to incorrect protective knowledge.

Each dimension is computed by averaging the encoded scores of relevant survey questions. The $\mathrm{A}_{i, k}^A$ factor measures familiarity-related concerns, while $\mathrm{B}_{i, k}^R$ captures the frequency of risk-enhancing behaviors, using scores derived from questions about daily or infrequent online activities. The Trust factor ($\mathrm{P}_{i, k}^C$) and Impulsivity factor ($\mathrm{P}_{i, k}^I$) are evaluated using responses about personal traits and decision-making tendencies, respectively. Encounters with cyber threats ($\mathrm{E}_{i, k}^E$) and responses to incidents ($\mathrm{E}_{i, k}^R$) are scored based on the frequency and severity of past experiences, including financial losses and emotional distress.

The final IVI score was computed as a weighted average of these factors by $\mathrm{IVI} = \frac{1}{7} \big(\mathrm{A}_{i, k}^A + \mathrm{A}_{i, k}^K + \mathrm{B}_{i, k}^R + \mathrm{P}_{i, k}^C + \mathrm{P}_{i, k}^I + \mathrm{E}_{i, k}^E + \mathrm{E}_{i, k}^R\big)$.  This comprehensive index quantifies individual vulnerability factors based on familiarity, behavior, experience, and personality traits. This approach captures multidimensional aspects of vulnerability across participants.

\paragraph{\bf Modeling the Attack Severity Index (ASI).}  ASI is derived from the iPoll dataset by encoding and aggregating survey responses across key dimensions: Frequency, Consequence, and Sophistication. Survey questions were mapped to numerical values using predefined encoding schemes, assigning higher scores to responses indicating greater risk or vulnerability. 

Frequency ($\mathrm{F}_{i, k}$) was calculated as the sum of responses from questions measuring the prevalence of cyber events, with a score of 5 indicating frequent occurrences. Consequence ($\mathrm{C}_{i, k}$) aggregated scores from questions reflecting emotional, physical, or financial impacts, capturing outcome severity. Sophistication ($\mathrm{S}_{i, k}$) represented situational plausibility by summing scores from questions assessing the perceived legitimacy and sophistication of threats.

The ASI was computed as a weighted combination of these three components, with equal weights assigned to ensure a balanced vulnerability assessment. This comprehensive metric captures event frequency, outcome severity, and threat sophistication, offering insights into individual and systemic vulnerabilities. The encoded dataset facilitates further analysis across demographic and behavioral factors, enabling nuanced comparisons within the surveyed population.

The detailed feature extraction process for the IVI and ASI from the iPoll dataset is provided in Tables~\ref{tab:ivi-calculation} and \ref{tab:asi_calculation} in the appendices, respectively.

\subsection{SCVI Using the Reddit Scam Reports}

The Reddit Scam Reports dataset comprises user-generated posts from the subreddit r/scams, where individuals share experiences with scams. Spanning data from 2016 to 2023 (via the Pushshift API) and additional records from October to November 2024, the dataset focuses on scam reports for analysis. From an initial 5,000 observations, 450 scam reports were selected, ensuring equal representation across years.

Preprocessing steps included slang replacement, URL removal, and stripping mentions and hashtags. Text normalization involved converting to lowercase, expanding contractions, reducing elongated words, and removing non-alphanumeric characters while retaining punctuations. This cleaned dataset supports effective annotation of scam types and success metrics (successful or unsuccessful scams), providing a solid foundation for modeling and analyzing scam dynamics.

\paragraph{\bf Modeling the Individual Vulnerability Index (IVI).}  The SCVI assesses individual vulnerabilities through key dimensions that capture user-specific factors influencing susceptibility to cyber threats. These dimensions offer a holistic view of personal risk, encompassing awareness, behaviors, psychological attributes, and prior experiences, collectively termed IVI. Below, we outline the core components of IVI:
\begin{itemize}
    \item \textbf{Individual Lack of Awareness and Knowledge ($\mathrm{A}$):} This measure evaluates users' exposure to scam-prevention strategies via fine-grained annotations of scam types and success. Participation in virtual communities improves well-being and knowledge sharing, reducing scam susceptibility~\cite{ayachi2021virtual}. Frequent engagement with scam discussions fosters familiarity with scams and avoidance methods, lowering vulnerability.

    \item \textbf{Behavioral Traits ($\mathrm{B}$):} Behavioral vulnerability is modeled by analyzing user interaction frequency, such as posting behavior and linguistic markers. Frequent engagement with scam-related discussions, reflected in posts or reports, may indicate risk-taking or impulsive behavior linked to scam susceptibility~\cite{herman2018risk}. Tools like Linguistic Inquiry and Word Count (LIWC)~\cite{liwc22} assess markers such as clout scores, perceptual language, and informal markers~\cite{tausczik2010psychological}. For example, low clout scores suggest low confidence, while reliance on sensory descriptions indicates vulnerability to scams leveraging fake credibility cues. Behavioral traits like low self-confidence and over-reliance on perceptual processes significantly affect scam susceptibility~\cite{norris2019psychology}. Specific language patterns in online communities also influence susceptibility to scams and manipulative content \cite{ta2022inclusive}, while skepticism, reflected in negations ("no," "not," "never"), helps resist misinformation and identify scams~\cite{wright2020many}.

    \item \textbf{Psychological Factors ($\mathrm{P}$):} Analytical thinking, emotional states, and personality traits significantly influence scam susceptibility. LIWC outputs, such as "affect" scores for emotional language and "cognitive processes" scores for critical thinking, provide insights into these dimensions. Strong analytical and cognitive skills enhance scam resistance \cite{hruschka2023learning}, while better cognitive abilities improve financial decision-making \cite{gamble2015aging}. Emotional states like anxiety and neuroticism impair decision-making, increasing susceptibility \cite{boyle2022degraded, cho2016effect}, with stress further exacerbating vulnerability \cite{norris2021personality}. Credulity, rather than general trust, is a specific fraud risk factor among older adults \cite{shao2019credulity}. A psychological vulnerability score, derived using LIWC, identifies potential risks through markers like high emotional language, excessive exclamations, or overly positive tones ("posemo"), while skepticism and reflective thinking reduce vulnerability.

    \item \textbf{Experiences ($\mathrm{E}$):} Prior exposure to scams is evaluated through annotations of posts detailing personal experiences, such as financial losses or emotional distress. This includes identifying whether users were scam victims. Such exposure may increase awareness or reinforce susceptibility~\cite{houtti2024survey, sheng2010falls}.

\end{itemize}

\paragraph{\bf Modeling the Attack Severity Index (ASI).}  ASI is modeled by annotating scam reports for two critical attributes: {\em scam type} and {\em scam success}. The annotation process is conducted by two human annotators, with additional refinement using OpenAI’s API. Scam types include \textit{phishing}, \textit{investment scams}, \textit{lottery scams}, \textit{tech support scams}, \textit{romance scams}, \textit{online shopping scams}, \textit{job scams}, and \textit{undetected}. Scam success is represented as a binary variable (0 or 1). The ASI consists of three key components:
\begin{itemize}
    \item \textbf{Frequency ($\mathrm{F}$):} The frequency component quantifies the number of reported incidents for each scam type, providing insights into the prevalence of various scams~\cite{houtti2024survey}.

    \item \textbf{Consequence ($\mathrm{C}$):} The consequence of an attack assesses the impact of successful scams by analyzing financial losses~\cite{lwin2023supporting} and emotional distress~\cite{whitty2016online, safari2023emotion}. Annotated reports capture monetary losses, while NLP techniques like emotion recognition models evaluate emotional impacts by aggregating positive and negative emotions in scam reports.

    \item \textbf{Sophistication ($\mathrm{S}$):} The sophistication of an attack~\cite{darem2024beyond} measures the effectiveness of scams based on the proportion of successful scams relative to total reports within each type. Higher success rates indicate greater sophistication in deceiving victims, reflecting the severity of these scams.
\end{itemize}

\vspace{-5mm}

\section{Evaluation Setup}

\paragraph{\bf Sensitivity Analysis of Weighting Schemes on SCVI Scores Using iPoll and Reddit Data.} A sensitivity analysis was conducted to evaluate the effect of varying weighting schemes on SCVI scores using two datasets: iPoll and Reddit scam reports. The analysis focused on the IVI components (\(w_A, w_B, w_P, w_E\)) and ASI components (\(w_F, w_C, w_S\)). The impact of each weight on the mean SCVI score and its variability was visualized and quantitatively assessed. This analysis provides insights into the relative importance of different vulnerability factors and helps identify the most influential components driving SCVI variations.

\paragraph{\bf Monte Carlo Simulation for Analyzing Weight Variability and Uncertainty in SCVI Scores.}  Monte Carlo simulations examined the uncertainty in SCVI scores by employing random sampling to model the effects of weight variability on the SCVI distribution. The plausible weight ranges for IVI (\(w_A, w_B, w_P, w_E\)) and ASI (\(w_F, w_C, w_S\)) were defined, ensuring that each index's weights sum to one. Over 10,000 iterations, SCVI scores were recalculated, and aggregated results were analyzed to identify key patterns and weight configurations. This approach allows for a robust evaluation of potential fluctuations in SCVI scores, offering a deeper understanding of their stability under different weighting scenarios.

\paragraph{\bf Evaluation of SCVI Against CVSS and SVI.}  The SCVI was evaluated against established indices, including the Common Vulnerability Scoring System (CVSS)~\cite{NISTCVSSMetrics} and the Social Vulnerability Index (SVI)~\cite{CDCSVI}. CVSS is a widely used framework for rating software vulnerabilities, prioritizing management based on metrics such as Base, Temporal, and Environmental factors. Similarly, the SCVI employs metric-based mappings to assess fraud victimization, incorporating elements like attack vector, complexity, user interaction, and impacts on confidentiality, integrity, and availability.

The SVI quantifies community-level vulnerabilities to disasters and public health emergencies by focusing on sociodemographic and infrastructural factors. The SCVI extends these considerations by integrating survey data on individual behaviors, demographics, and cyber threat exposure, offering a composite view of individual vulnerability.

The evaluation involved calculating SCVI metrics and comparing them with CVSS and SVI scores derived from the iPoll dataset. This approach enabled a comprehensive analysis of SCVI's capacity to integrate cyber-related behavioral data with sociodemographic factors, bridging established frameworks. Results are presented using tables to analyze significance, correlations, and outliers across demographic groups, including age, race-ethnicity, and gender.
\vspace{-3mm}
\section{Analyses of Evaluation Results}

SCVI was estimated using the iPoll and Reddit datasets and compared with two existing metrics: SVI and CVSS.

\subsection{Analyses of SCVI}

\begin{figure*}[ht]
    \centering
    \subfigure[Distributions of IVI factors]{
        \includegraphics[width=0.3\textwidth, height=0.2\textwidth]{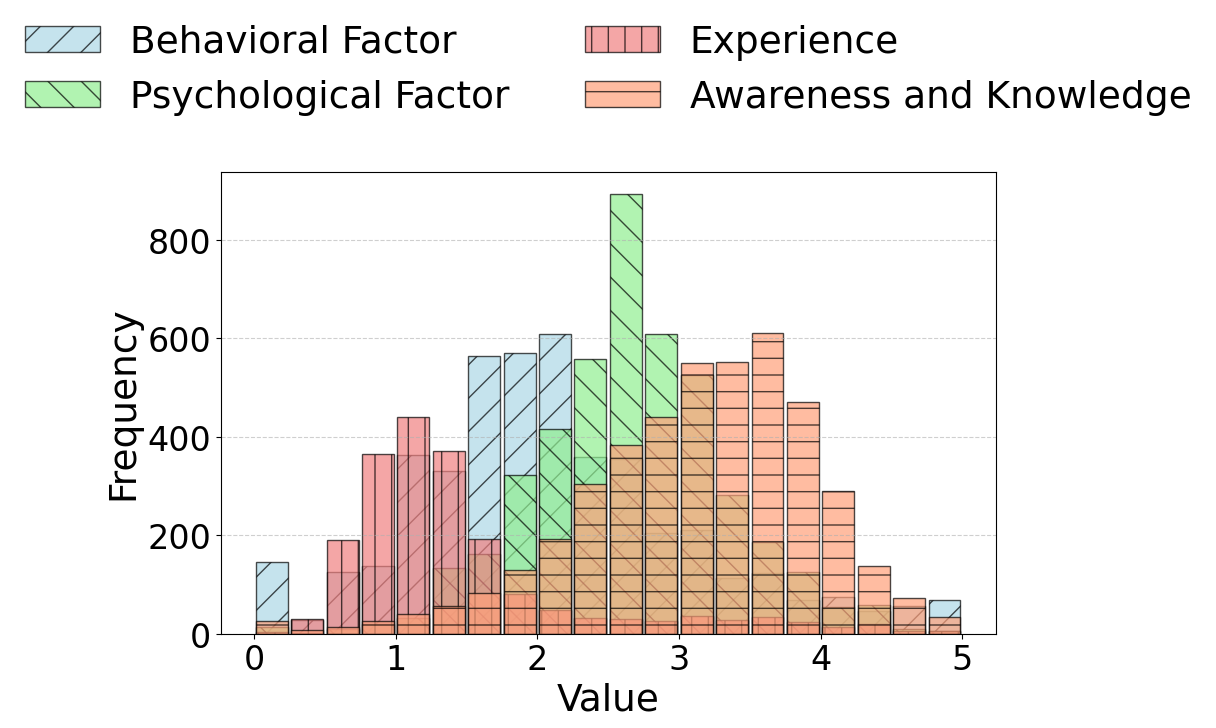}
        \label{fig:ipoll_ivi_factors}
    }
    \hspace{1mm}
    \subfigure[Distributions of IVI and ASI values]{
        \includegraphics[width=0.3\textwidth, height=0.2\textwidth]{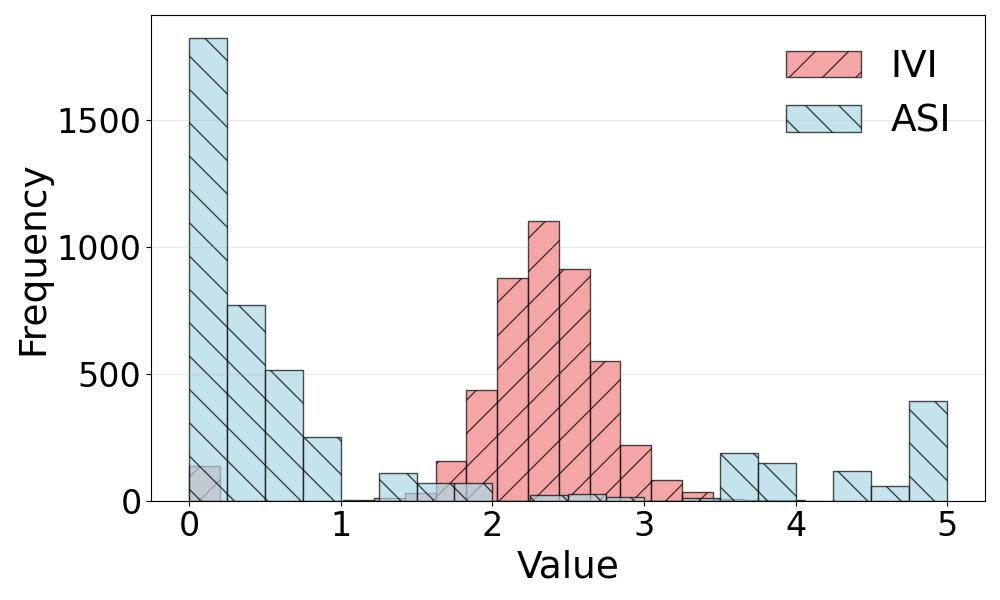}
        \label{fig:ipoll_ivi_asi}
    }
    \hspace{1mm}
    \subfigure[Distributions of ASI factors]{
        \includegraphics[width=0.3\textwidth, height=0.2\textwidth]{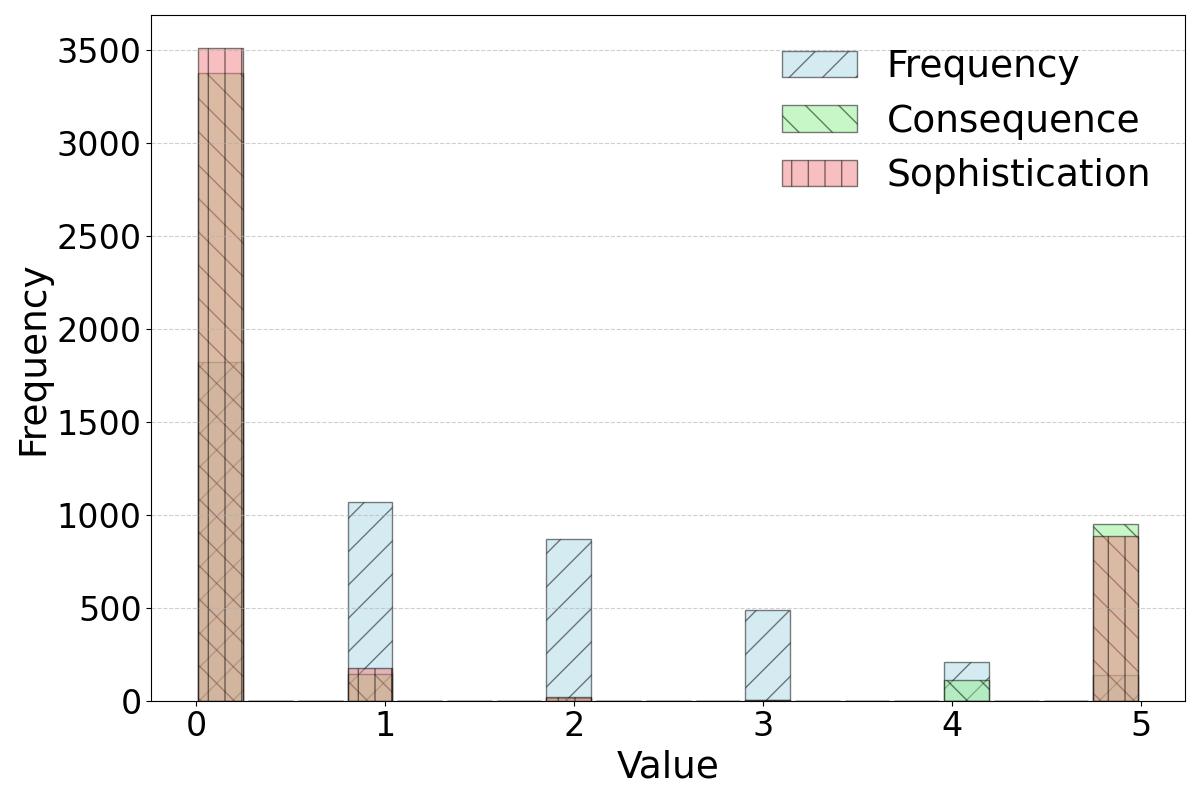}
        \label{fig:ipoll_asi_factors}
    }
    \caption{Comparison of IVI and ASI distributions and their contributing factors using the iPoll dataset. }
    \label{fig:combined-ipoll}
\end{figure*}

\paragraph{\bf Analysis of the iPoll Dataset.} The analysis of the iPoll dataset reveals significant insights into user vulnerabilities and attack severity. Figure~\ref{fig:ipoll_ivi_factors} illustrates the distribution of the IVI factors. The \textit{Behavioral} factor exhibits a concentration of values around 1, indicating that most users engage minimally in behaviors that might increase their susceptibility to scams. However, the \textit{Psychological} factor shows a more even distribution, with most users scoring between 2 and 4, reflecting moderate psychological attributes influencing their vulnerability. The \textit{Experience} factor is heavily skewed towards lower values, suggesting that many users have little to no prior exposure to scams. Meanwhile, the \textit{Awareness and Knowledge} factor is widespread, with many users demonstrating low to moderate awareness of scams.

\begin{figure*}[ht]
    \centering
    \subfigure[Distributions of IVI factors]{
        \includegraphics[width=0.3\textwidth, height=0.25\textwidth]{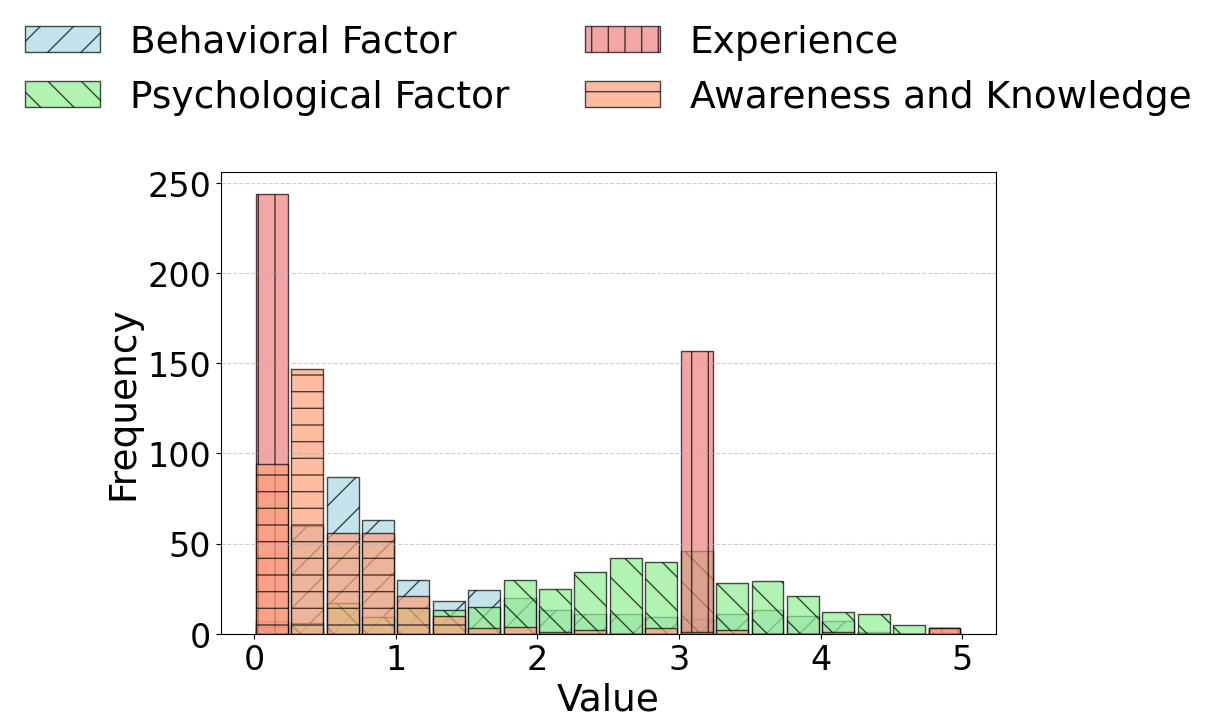}
        \label{fig:ivi_factors}
    }
    \hspace{1mm}
    \subfigure[Distributions of IVI and ASI values]{
        \includegraphics[width=0.3\textwidth, height=0.2\textwidth]{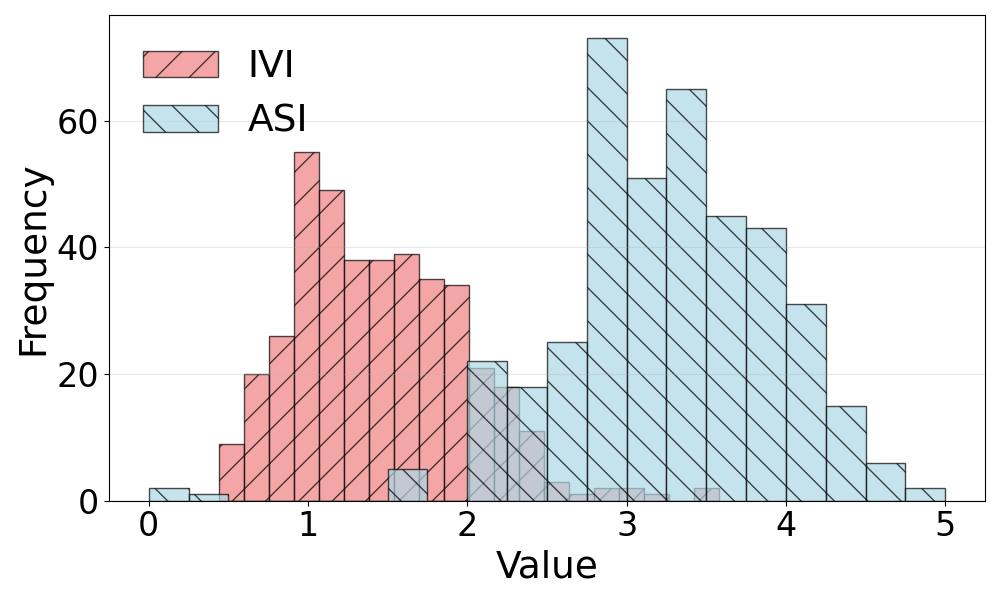}
        \label{fig:ivi_asi}
    }
    \hspace{1mm}
    \subfigure[Distributions of ASI factors]{
    \includegraphics[width=0.3\textwidth, height=0.2\textwidth]{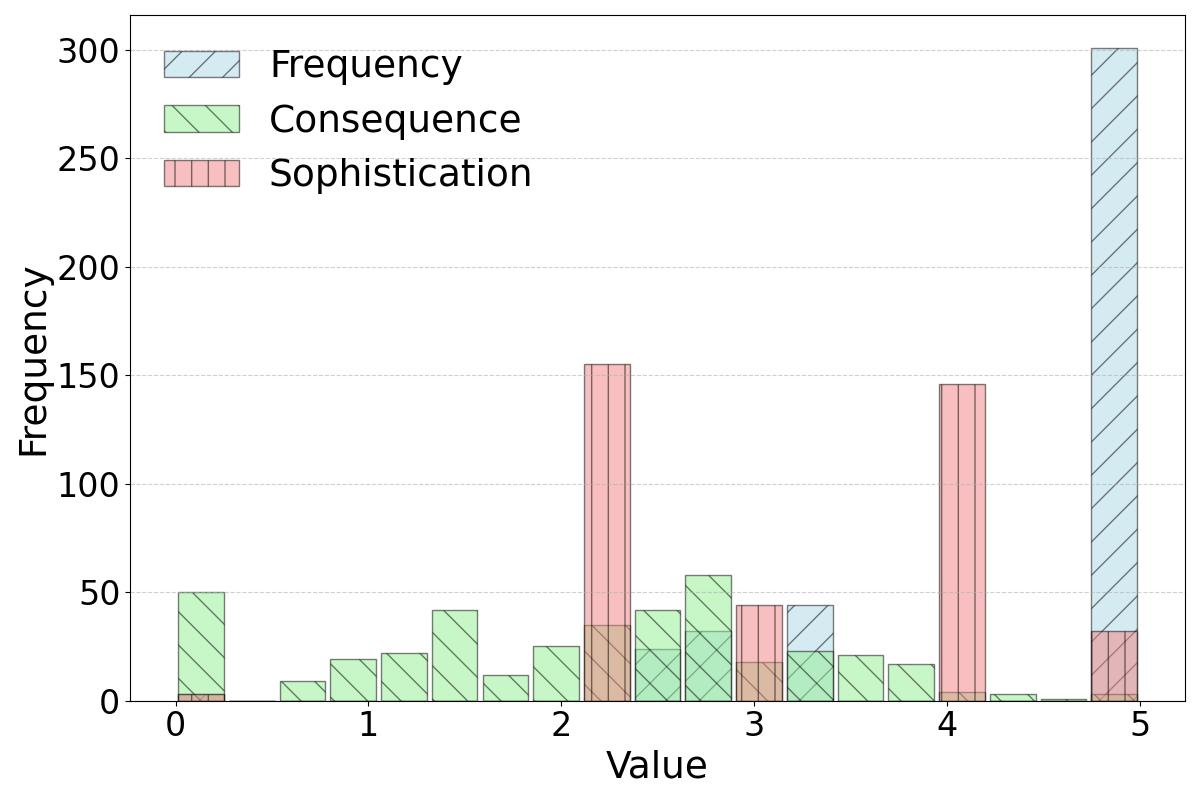}
        \label{fig:asi_factors}
    }
    \caption{Comparison of IVI and ASI distributions and their contributing factors using the Reddit dataset.}
    \label{fig:combined-reddit}
\end{figure*}

Figure~\ref{fig:ivi_asi} compares the IVI and ASI. The IVI distribution is between 2 and 3, reflecting moderate vulnerability levels among most users. In contrast, the ASI distribution is heavily skewed towards lower values, with a substantial frequency at 0, indicating that many users face attacks with minimal severity. However, a smaller yet significant group experiences higher ASI values around 4 and 5, highlighting the presence of severe attack cases. This disparity underscores the need to address user vulnerabilities and the impact of high-severity attacks to ensure comprehensive protection.

The factors contributing to the ASI provide additional insights, as shown in Figure~\ref{fig:ipoll_asi_factors}. The \textit{Frequency} factor reveals a bimodal distribution, with peaks near 0 and 5, indicating that users either rarely or frequently encounter scams. The \textit{Consequence} factor is concentrated around moderate values, reflecting that the overall impact of scams is neither negligible nor catastrophic for most users. The \textit{Sophistication} factor shows a diverse distribution, with peaks near both ends of the scale. This suggests a wide variation in the perceived authenticity and persuasiveness of scam attempts, with some being highly convincing while others are easily recognized.

In conclusion, the analysis of the iPoll dataset demonstrates a consistent pattern of low user vulnerability paired with high attack severity. The distribution of IVI and ASI factors suggests that while users may not frequently engage in risky behaviors, they remain vulnerable to severe and highly convincing scams. This highlights the need for targeted educational initiatives and awareness campaigns to reduce the impact of scams on vulnerable populations.

\paragraph{\bf Analysis of Reddit Dataset.}  The distribution of the Individual Vulnerability Index (IVI) factors provides insights into user vulnerability, as shown in Figure~\ref{fig:ivi_factors}. The \textit{Behavioral} factor displays a strong skew towards lower values, indicating that most users exhibit low engagement in behaviors that might increase their susceptibility to scams. In contrast, the \textit{Psychological} factor demonstrates a more balanced distribution, with most users scoring between 2 and 4, suggesting moderate psychological attributes contributing to vulnerability. The \textit{Experience} factor is concentrated at discrete points, with a significant number of users reporting no prior scam encounters, while a smaller group shows high values, reflecting substantial past interactions with scams. Lastly, the \textit{Awareness and Knowledge} factor is heavily skewed towards lower values, highlighting a general lack of awareness and knowledge about scams among users.

A comparison of the IVI and the Attack Severity Index (ASI) reveals a clear distinction in their distributions, as shown in Figure~\ref{fig:ivi_asi}. The IVI primarily concentrates between 1 and 2, reflecting low to moderate vulnerability levels for most users. Conversely, the ASI distribution is shifted towards higher values, with the majority of users scoring between 3 and 4. This suggests that while users might exhibit relatively low vulnerability, the severity of the attacks they encounter is notably higher. This disparity underscores the need for targeted measures to bridge the gap between user vulnerability and the impact of the scams they experience.

The factors contributing to the ASI provide further insights into the nature of scams, as illustrated in Figure~\ref{fig:asi_factors}. The \textit{Frequency} factor is highly polarized, with peaks near 0 and 5, suggesting that users experience either very frequent or infrequent attacks. The \textit{Consequence} factor has a moderate peak around 3, indicating that the impact of scams is generally moderate for most users. However, the \textit{Sophistication} factor, which measures the effectiveness of scams, displays a bimodal distribution with peaks near 2 and 4. This highlights varying levels of success and persuasiveness in scam attempts, with some being highly convincing while others are less so.

In conclusion, the analysis highlights a critical mismatch between user vulnerability and the severity of attacks. While most users exhibit low IVI scores, the high ASI values indicate that the attacks they encounter are often severe and impactful. Addressing this disparity requires enhancing user awareness and psychological resilience, particularly in recognizing and mitigating high-severity scams. These findings provide a foundation for designing effective strategies to reduce the impact of scams on vulnerable populations.

\subsection{Sensitivity Analysis of IVI and ASI Components on SCVI Variability}

\begin{figure*}[ht]
    \centering
    \subfigure[Varying the weights of the IVI factors]{
        \includegraphics[width=0.45\textwidth]{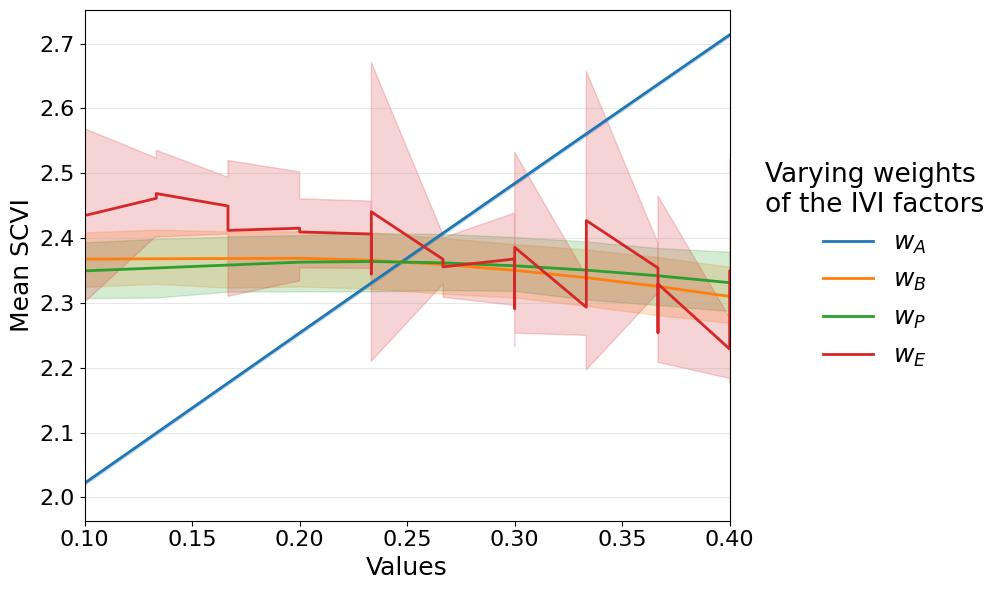}
        \label{fig:reddit_ivi_sensitivity}
    }
    \hfill
    \subfigure[Varying the weights of the ASI factors]{
        \includegraphics[width=0.45\textwidth]{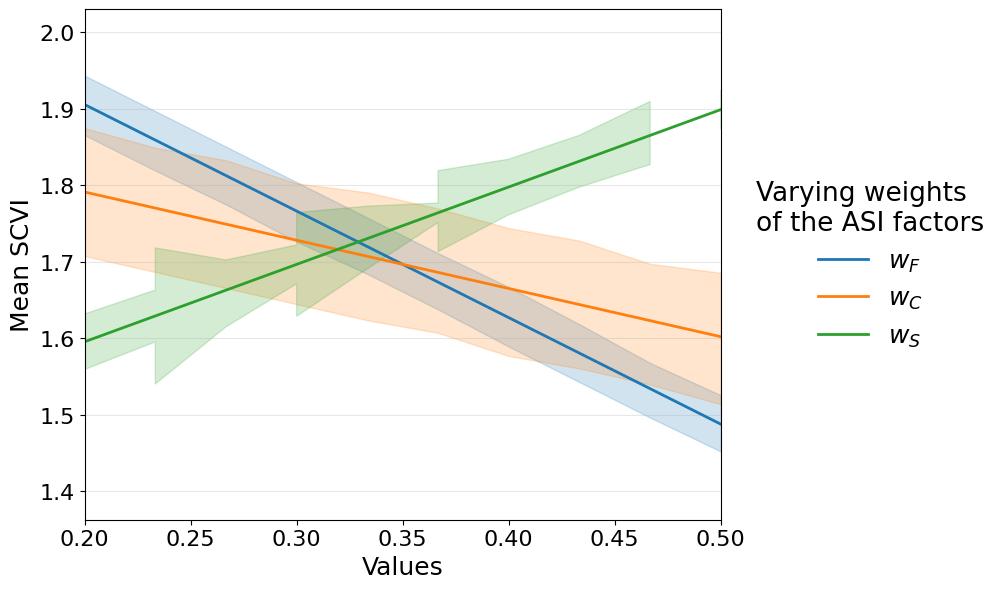}
        \label{fig:reddit_asi_sensitivity}
    }
    \caption{Sensitivity analysis of the Reddit dataset for IVI and ASI factors. Note that $w_A$, $w_B$, $w_P$, and $w_E$ are the weights for `Awareness,' `Behavioral,' `Psychological,' and `Experience' factors in the individual vulnerability index (IVI) correspondingly. $w_F$, $w_C$, and $W_S$ refer to `Frequency,' `Consequence,' and `Sophistication' in the Attack Security Index (ASI), respectively.  }
    \label{fig:reddit_sensitivity}
\end{figure*}

The sensitivity analysis on the ipoll dataset reveals key trends in the contributions of IVI and ASI components to SCVI variability, as shown in Figure~\ref{fig:reddit_sensitivity}. In the IVI, the {\it Awareness} factor (\(w_A\)) consistently increased SCVI scores, highlighting the exacerbating effect of lack of awareness. The {\it Experience} factor (\(w_E\)) showed a decreasing trend with fluctuations, indicating interactions with other components. The {\it Psychological} factor (\(w_P\)) strongly correlated with increased SCVI, while the {\it Behavioral} factor (\(w_B\)) had minimal positive influence, reflecting limited impact in the Reddit context.

Regarding ASI components within the ipoll dataset, the {\it Frequency} factor (\(w_F\)) showed a strong negative correlation with SCVI, indicating that lower frequencies of cyber-attacks increase vulnerability. The {\it Consequence} factor (\(w_C\)) consistently decreased SCVI scores as its weight increased, reflecting the mitigating effect of preparedness against high-consequence events. The {\it Sophistication} factor (\(w_S\)) displayed a mild positive trend with variability, suggesting its impact on SCVI depends on interactions with other factors.

\begin{figure*}[ht]
    \centering
    \subfigure[Varying the weights of the IVI factors]{
        \includegraphics[width=0.45\textwidth]{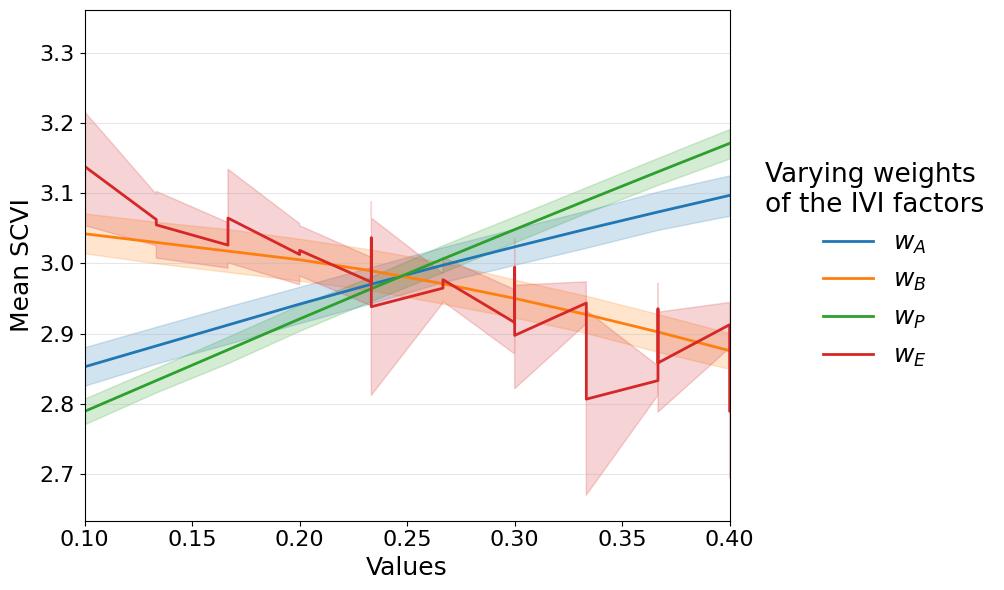}
        \label{fig:ipoll_ivi_sensitivity}
    }
    \hfill
    \subfigure[Varying the weights of the ASI factors]{
        \includegraphics[width=0.45\textwidth]{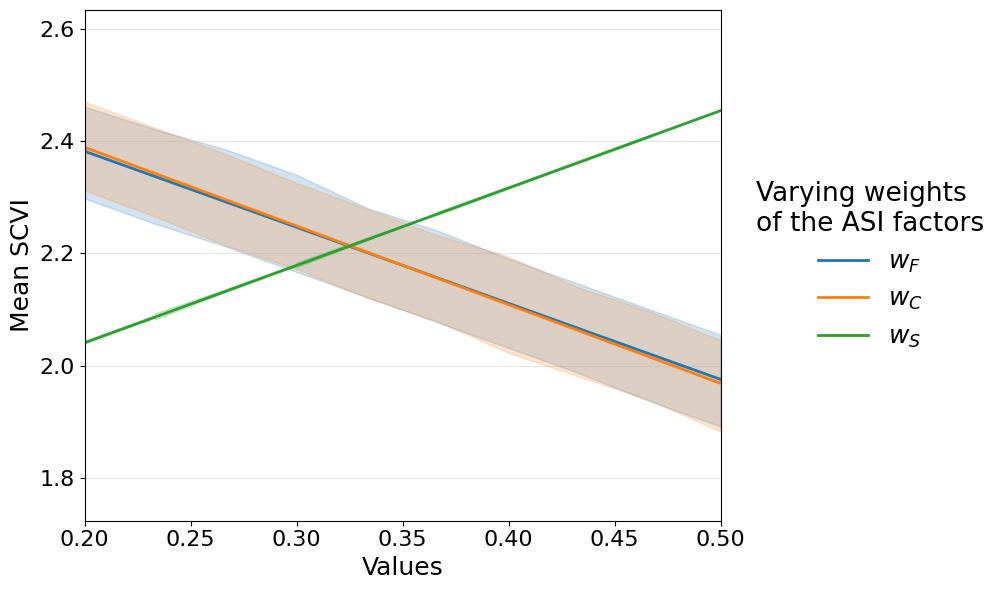}
        \label{fig:ipoll_asi_sensitivity}
    }
    \caption{Sensitivity analysis of the iPoll dataset for IVI and ASI factors. Note that $w_A$, $w_B$, $w_P$, and $w_E$ are the weights for `Awareness,' `Behavioral,' `Psychological,' and `Experience' factors in the individual vulnerability index (IVI) correspondingly. $w_F$, $w_C$, and $w_S$ refer to `Frequency,' `Consequence,' and `Sophistication' in the Attack Security Index (ASI), respectively.}
    \label{fig:ipoll_sensitivity}
\end{figure*}

The analysis of the Reddit dataset mirrored ipoll findings mostly but revealed differences in factor impacts, as in Figure~\ref{fig:ipoll_sensitivity}. The strong exacerbating effects of higher weights in {\it Awareness (\(w_A\)) and mitigating effects of Experience (\(w_E\))} on SCVI were consistent across both datasets. In contrast, the {\it Psychological} factor (\(w_P\)) exhibited a slight negative driver of SCVI. The {\it Behavioral} factor (\(w_B\)) showed a slight negative correlation with SCVI in the Reddit dataset, in comparison with the ipoll analysis. For the ASI components, Frequency (\(w_F\)) was again a major contributor to SCVI in both datasets, though its impact was stronger in the Reddit data. The {\it Consequence} factor (\(w_C\)) similarly reduced SCVI across both datasets but had a more pronounced effect in the Reddit dataset. The {\it Sophistication} factor (\(w_S\)) had a more consistent and significant positive impact  on SCVI in the iPoll dataset compared to its less stable influence in the Reddit dataset.

These analyses underscore the robust yet context-sensitive nature of SCVI components across different datasets. The universal impacts of  Awareness, Experience, Consequence and Frequency on SCVI were evident. Nonetheless, variations in the magnitude and direction of relationships for Psychological, Behavioral and Sophistication factors across datasets highlight the influence of data-specific attributes. Further investigations into additional datasets and varying contexts can enhance the applicability and robustness of SCVI as a comprehensive metric for assessing cyber vulnerabilities.  We leave this for our future research.

\subsection{Effect of Weight Variability in SCVI}

\begin{figure*}[ht]
  \centering
    
\setcounter{subfigure}{0}
  \subfigure[Monte Carlo Analysis of SCVI Components in the iPoll Dataset: Primary and Secondary Peaks.]{    \includegraphics[width=0.45\textwidth, height=0.25\textwidth]{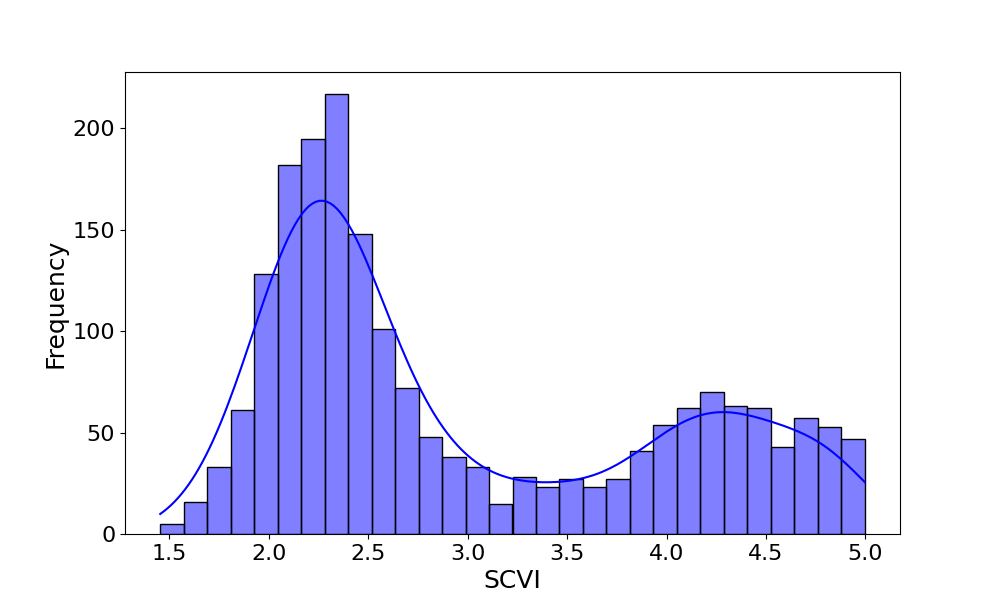} \label{fig:montecarlo_ipoll}}
    \subfigure[Monte Carlo Analysis of SCVI Components in the Reddit Dataset: Low and High SCVI Outliers.]{\includegraphics[width=0.45\textwidth, height=0.25\textwidth]{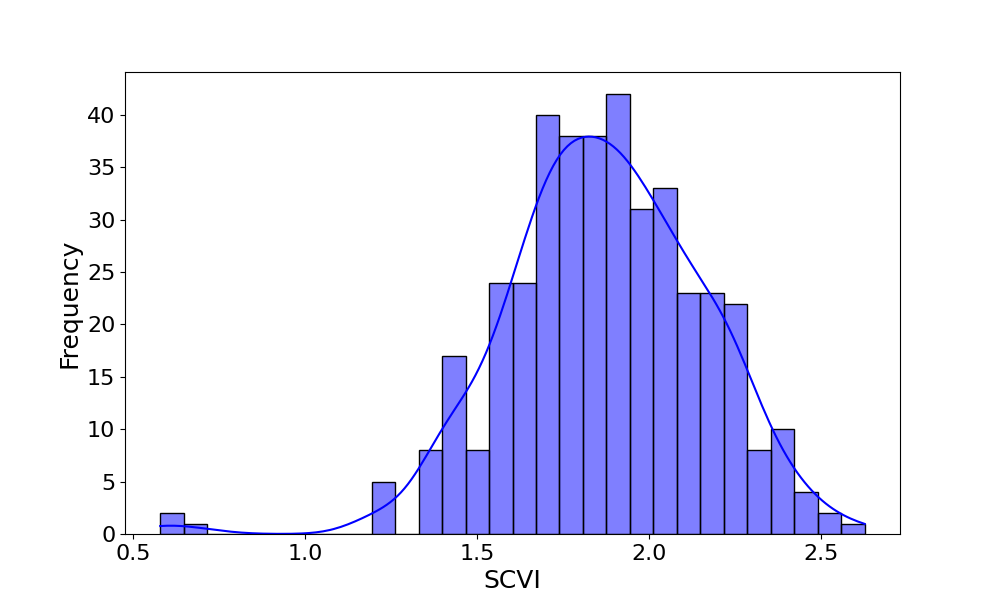}
    \label{fig:montecarlo_reddit}}
    \caption{Monte Carlo analysis of SCVI components in the iPoll and Reddit datasets.}
\end{figure*}

The SCVI values and corresponding weight configurations were aggregated across multiple iterations to identify key patterns, as shown in Figures~\ref{fig:montecarlo_ipoll} and~\ref{fig:montecarlo_reddit}. The results highlighted two primary peaks shown in Figure~\ref{fig:montecarlo_ipoll} for the iPoll dataset:

\textbf{Primary Peak (Group 1)}: SCVI was predominantly driven by Experience (\( w_E \)) in the Individual Vulnerability Index (IVI) and Sophistication (\( W_S \)) in the Attack Severity Index (ASI). Experience made the largest contribution to IVI, with a mean value of 0.421 and a standard deviation of 0.084. Similarly, Sophistication dominated ASI contributions, with a mean value of 0.447 and a standard deviation of 0.095. These findings emphasize the importance of systemic Sophistication and individual experience in influencing SCVI scores.

\textbf{Secondary Peak (Group 2)}: SCVI values were primarily influenced by Awareness (\( w_A \)) and Psychological Factors (\( w_P \)) in IVI, as well as Frequency (\( w_F \)) in ASI. Awareness emerged as a significant IVI contributor (Mean: 0.372, Std: 0.026), while Frequency was the dominant ASI factor (Mean: 0.436, Std: 0.033). In this group, Sophistication and Experience played a minimal role, with SCVI relying more on attack frequency, consequences, and psychological factors.

We find the following observation in Figure~\ref{fig:montecarlo_reddit}, which analyzed the Reddit dataset.

\textbf{Low SCVI Outliers}: SCVI values exhibited low variability, ranging from 1.039 to 1.098, indicating consistency. Behavioral Factors (\( w_B \)) emerged as the dominant contributor, ranging from 0.33 to 0.37, with Experience (\( w_E \)) also playing a significant role in certain cases. $\alpha$ values ranged between 0.562 and 0.593, slightly favoring IVI, while $\beta$ values from 0.406 to 0.437 represented a balanced ASI contribution. This indicates that user behaviors, such as risk-taking tendencies and prior experiences, significantly influence SCVI, while the overall stability of the metric reflects a robust model for assessing vulnerabilities.

\textbf{High SCVI Outliers}: SCVI values showed greater variability, ranging from 2.182 to 2.211. Experience (\( w_E \)) and Frequency (\( w_F \)) were the most influential factors, consistently exceeding 0.45, highlighting the importance of users' prior exposure to scams and the prevalence of cyberattack attempts in shaping vulnerabilities. and $\alpha$ values between 0.403 and 0.419 suggested a balanced primary contribution from IVI components, while and $\beta$ values between 0.580 and 0.596 indicated a stronger impact from ASI components. This distribution suggests that while individual vulnerability factors contribute significantly, the attack-specific characteristics, particularly frequency, dominate overall SCVI scores.

\textbf{Comparative Analysis}: Both datasets highlight the dominant roles of Experience (\( w_E \)) and Frequency (\( w_F \)) in driving SCVI scores. However, the Reddit dataset emphasized Behavioral Factors (\( w_B \)) in low SCVI cases and a more pronounced influence of Frequency in high SCVI cases, suggesting that user behaviors are more critical in reducing vulnerability, while the prevalence of cyberattacks becomes a stronger driver of high SCVI scores. In contrast, the iPoll dataset demonstrated stronger contributions from Sophistication (\( W_S \)) and Awareness (\( w_A \)), reflecting the importance of user Awareness and the complexity of attacks in shaping vulnerabilities. This comparison underscores the contextual differences between datasets, highlighting the varying impacts of individual behaviors and attack characteristics on SCVI.

Summarizing the insights above, while SCVI components exhibit universal trends, their relative contributions differ significantly across datasets, reflecting the contextual characteristics of the data. For instance, the iPoll dataset highlights the dominant roles of Sophistication (\( W_S \)) and Awareness (\( w_A \)), emphasizing the importance of attack complexity and user awareness in shaping vulnerabilities. Conversely, the Reddit dataset underscores the critical influence of Behavioral Factors (\( w_B \)) in low SCVI cases and Frequency (\( w_F \)) in high SCVI cases, illustrating the varying impacts of user behaviors and cyberattack prevalence. 

This variability highlights that tailoring SCVI metrics is critical to the specific attributes of the analyzed dataset. Future research could delve deeper into these dynamics to enhance SCVI's adaptability and robustness, enabling more accurate assessments of vulnerabilities across diverse environments.

\subsection{Comparative Analysis with SVI and CVSS Across Demographic Groups}
\begin{figure}[ht]
    \centering
    \includegraphics[width=0.9\columnwidth]{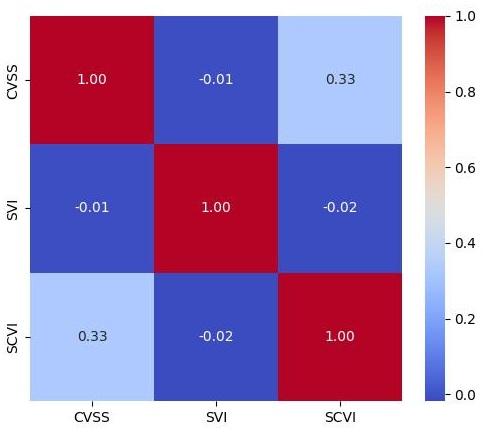}
    \caption{Spearman correlation heatmap across SCVI, SVI, and CVSS metrics.}
    \label{fig:spearman_correlation}
\end{figure}

The SCVI metric effectively captures individual-level vulnerabilities in social cyber contexts, surpassing SVI and CVSS through its integration of socio-demographic, behavioral, and cyber-specific dimensions. As shown in Figure~\ref{fig:spearman_correlation}, correlation analysis reveals a moderate positive correlation with CVSS (Spearman = 0.33, p = 0.0), indicating alignment in identifying technological vulnerabilities. In contrast, SCVI exhibits a weaker correlation with SVI (-0.01, p = 0.4836), reflecting SVI's broader socio-environmental focus, which lacks specificity in cyber-related risks.

SCVI complements CVSS and SVI by incorporating dimensions that address individual vulnerabilities in socio-cyber contexts. Its alignment with CVSS and divergence from SVI highlight its ability to integrate social and behavioral factors, making it a vital tool for inclusive cybersecurity strategies.

Analysis of trends across gender, race-ethnicity, and age groups reveals consistent patterns in vulnerability metrics. Table~\ref{tab:gender} shows that the male group has the highest SVI, CVSS, and SCVI values at 2.57, 3.52, and 3.47, respectively. The SVI difference between male and female groups is relatively minimal (2.36\%), while CVSS and SCVI show significantly larger gaps at 9.84\% and 16.88\%, potentially due to exposure to riskier technological environments.

As shown in Table~\ref{tab:race_ethnicity}, among racial and ethnic groups, the White, non-Hispanic group has the lowest values across all indices, with SVI at 2.25 and CVSS at 3.02. Conversely, the Hispanic group records the highest values, with SVI, CVSS, and SCVI at 3.72, 3.79, and 4.06, respectively. SCVI generally exceeds CVSS, which in turn exceeds SVI. These findings suggest that targeted cybersecurity programs for Hispanic groups could address their higher vulnerability. In contrast, the lower scores for White, non-Hispanic groups may reflect better access to resources, education, or programs.

Holistically, the three columns in Table~\ref{tab:race_ethnicity} showcase a trend of SVI scores almost always being lower than that of CVSS or SCVI. In fact, the only two groups where SVI is not the lowest is: black, non-Hispanic (by 0.3\%) and other, non-Hispanic (by 10\%) race-ethnicity groups in the table. No group displayed SVI having the highest vulnerability index.

\begin{table}[t]
    \centering
    \small
    \caption{\sc \centering Distribution of SVI, CVSS, and SCVI metrics by Gender}
    \label{tab:gender}
    \begin{tabular}{|l|c|c|c|}
        \hline
        Gender  & SVI  & CVSS & SCVI \\ \hline
        Female  & 2.51 & 3.19 & 2.93 \\ \hline
        Male    & 2.57 & 3.52 & 3.47 \\ \hline
    \end{tabular}
\end{table}

\begin{table}[t]
    \centering
    \small 
    \caption{\sc \centering Distribution of SVI, CVSS, and SCVI metrics by Race-Ethnicity}
    \label{tab:race_ethnicity}
    \begin{tabular}{|c|c|c|c|}
        \hline
        Race-Ethnicity          & SVI  & CVSS & SCVI \\ \hline
        White, non-Hispanic     & 2.25 & 3.31 & 3.02 \\ \hline
        Black, non-Hispanic     & 3.34 & 3.34 & 3.40 \\ \hline
        Other, non-Hispanic     & 3.01 & 3.74 & 3.33 \\ \hline
        2+, non-Hispanic        & 2.89 & 3.32 & 3.14 \\ \hline
        Asian, non-Hispanic     & 2.90 & 3.32 & 3.02 \\ \hline
        Hispanic                & 3.72 & 4.06 & 3.79 \\ \hline
    \end{tabular}
\end{table}

\begin{table}[t]
    \centering
    \small 
    \caption{\sc \centering Distribution of SVI, CVSS, and SCVI metrics by Age Groups}
    \label{tab:age_groups}
    \begin{tabular}{|c|c|c|c|}
        \hline
        Age Group & SVI  & CVSS & SCVI \\ \hline
        18-24     & 3.85 & 4.69 & 4.14 \\ \hline
        25-29     & 3.11 & 3.59 & 3.50 \\ \hline
        30-44     & 3.31 & 3.33 & 3.11 \\ \hline
        45-49     & 2.74 & 3.82 & 3.45 \\ \hline
        50-54     & 2.42 & 3.81 & 3.45 \\ \hline
        55-64     & 2.25 & 3.24 & 2.95 \\ \hline
        65+       & 2.12 & 2.82 & 2.63 \\ \hline
    \end{tabular}
\end{table}

In age group analysis, presented in Table~\ref{tab:age_groups}, younger demographics (18-24 and 25-29) are overall more vulnerable than older age groups with SCVI being the highest. This indicates a stronger susceptibility to social cyber vulnerabilities possibly due to higher technology usage. As age increases, all metrics notably decrease with middle-aged and older-aged groups’ CVSS scores leading the other two metrics. Note that there is a slight increase in all metrics for the 45-49 and 50-54 age groups, perhaps indicating higher engagement but less awareness with technology and scams than neighboring age groups. SVI scores are also the lowest across all age groups.

These findings validate SCVI as a robust and comprehensive metric, surpassing traditional indices in capturing nuanced social cyber vulnerabilities. SCVI consistently exhibits the highest scores across demographic categories, emphasizing its utility in identifying risks associated with technological behaviors and social factors. This distinction highlights its potential as a comprehensive tool for addressing individual-level vulnerabilities, particularly in demographic groups with higher susceptibility to cyber threats.

\subsection{Regional Disparities in Cyber Vulnerability: Analysis of SCVI Across U.S. States Using the iPoll Dataset}

The SCVI calculated from the iPoll dataset analyses regional disparities in cyber vulnerability across the United States. The dataset includes mean SCVI scores, confidence intervals, and sample sizes for each state, providing critical insights into the geographical distribution of cyber threats and the effectiveness of local cybersecurity measures.

Data was visualized through a heatmap to illustrate the SCVI distribution, with color intensities adjusted based on sample sizes to reflect data reliability in Figure~\ref{fig:heatmap}. Table~\ref{tab:state_summary} in the appendices provides a detailed summary of the state-wise SCVI scores, sample sizes, and confidence intervals.

States like Alaska, Rhode Island, and Nevada, which exhibit higher mean SCVI scores and smaller sample sizes, suggest an elevated risk level potentially influenced by regional factors. The wide confidence intervals in these measurements indicate a significant uncertainty in the SCVI estimates, attributed to the limited data availability.

In contrast, populous states such as California, Texas, and New York demonstrate lower and more stable SCVI scores, suggesting either lower cyber vulnerability or more effective cybersecurity practices . The robustness of data from these states provides clearer insights into their cyber health landscapes. The variability in SCVI scores between states like Connecticut and Pennsylvania, despite similar sample sizes, underscores the impact of socio-economic or infrastructural factors on cyber vulnerability. This variability highlights the complex dynamics affecting cybersecurity across regions.These observations align with the Federal Trade Commission's (FTC) fraud reports in the 2020 FTC Data Book \cite{FTC2021}, which highlight similar regional trends in cyber fraud incidents and vulnerabilities.

The inverse relationship observed between the mean SCVI scores and sample sizes across several states suggests that smaller samples may capture specific regional extremes or anomalies not present in larger, more representative data sets. This observation necessitates cautious interpretation of data and, possibly, further investigation into these regions.

The analysis underscores the need for region-specific cyber vulnerability assessments and tailored cybersecurity policies addressing unique challenges across states. Enhancing data collection efforts and expanding sample sizes can improve the reliability of future SCVI assessments and develop more effective regional and national cybersecurity strategies.

\begin{figure}[t]
\centering
\includegraphics[width=0.48\textwidth]{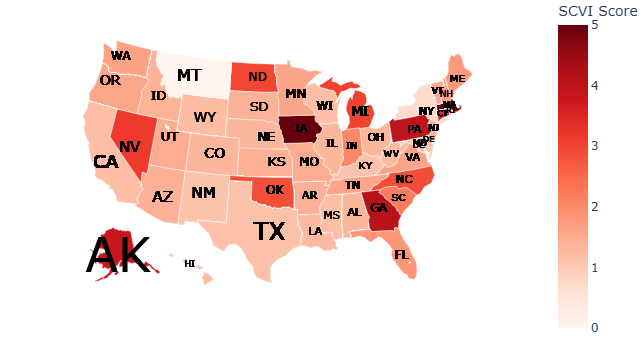}
\caption{Heatmap of the SCVI distribution across the United States, with color intensities reflecting sample sizes to indicate data reliability.}
\label{fig:heatmap}
\end{figure}

\section{Conclusions, Limitations, \& Future Work}

\subsection{Discussions of Key Findings}

The findings across both the iPoll and Reddit datasets reveal significant trends in user vulnerabilities and attack characteristics. Individuals often exhibit relatively low self-assessed vulnerability ({\it IVI}) yet face scams with high severity ({\it ASI}), highlighting a critical mismatch. This discrepancy underscores the need for targeted interventions to increase awareness and improve coping strategies, mainly because many users appear insufficiently prepared to handle high-impact attacks such as phishing, romance scams, and investment fraud despite engaging in minimal risk-taking behaviors.

The dimensions of {\it psychological factors} (e.g., impulsivity, credulity, emotional distress) and {\it experiences} (e.g., falling for previous scams) proved incredibly influential in shaping vulnerability scores. Users reporting higher impulsivity or emotional distress consistently showed elevated IVI values, emphasizing the interplay between personal disposition and cyber risk. This insight calls for intervention strategies to enhance psychological resilience to mitigate vulnerability.

A key strength of the {\it SCVI} is its integration of individual-level (IVI) and attack-level (ASI) metrics, surpassing traditional indices such as {\it CVSS}, which prioritizes technical vulnerabilities, and {\it SVI}, which emphasizes broad social and infrastructural vulnerabilities. SCVI's moderate positive correlation with CVSS ($r=0.33$) highlights its alignment with technological measures. In contrast, its minimal correlation with SVI ($r \approx -0.01$) demonstrates its unique capability to address cyber-specific behavioral and psychological dimensions. This refined granularity facilitates more nuanced intervention strategies, such as identifying scam types (e.g., tech support vs. investment) that present the most realistic or effective methods of victimization.

{\bf Sensitivity and weight variability analyses} revealed that {\it Awareness} ($w_A$) consistently increased SCVI scores, emphasizing its exacerbating effect, while {\it Experience} ($w_E$) generally mitigated SCVI with some fluctuations due to component interactions. The {\it Psychological} factor ($w_P$) showed a strong positive correlation with SCVI in ipoll but a slight negative effect in Reddit. Contextual differences were notable: lower scam frequencies in Reddit increased SCVI, whereas in iPoll, {\it Sophistication} ($w_S$) consistently reduced SCVI by shaping scam perceptions. These findings highlight the need for tailored interventions focusing on awareness, impulsive behaviors, and preparedness for high-severity scams.

{\bf Geographical analysis} using the iPoll dataset showed variations in SCVI scores influenced by regional factors and sample sizes. States with smaller sample sizes, such as Alaska and Rhode Island, often presented higher SCVI scores with wider confidence intervals, suggesting that localized scam dynamics or outlier incidents might skew risk perceptions. In contrast, populous states like CA and TX exhibited lower, more stable SCVI values, potentially reflecting more effective cybersecurity awareness efforts. These disparities underscore the need for region-specific cybersecurity policies, such as community-based digital literacy training and public awareness campaigns targeting prevalent scam types.

\subsection{Limitations of the Present Research}
Despite the robustness and novelty of the {\bf SCVI framework}, several limitations merit consideration.

{\bf First}, both the iPoll survey \cite{ipoll-dataset2020} and Reddit scam reports rely on self-reported data, which introduces {\bf risks of recall bias and under-reporting}. Users may understate their vulnerabilities or fail to accurately report certain scam experiences, potentially affecting the reliability of the findings. 

{\bf Second}, while the iPoll dataset strives to be nationally representative, certain demographics, such as older adults with limited Internet access or non-English speakers, may be underrepresented. Similarly, Reddit data are prone to {\bf self-selection bias}, favoring technologically savvy users, which could skew the results toward a specific subset of the population.

{\bf Third}, another limitation stems from {\bf sample size variability}. Some states exhibit very small sample sizes, resulting in wide confidence intervals and reducing the reliability of geographical analyses. This emphasizes enhancing data collection efforts across diverse regions to ensure more robust and representative findings. 

{\bf Forth}, the equal weighting of IVI and ASI components, applied uniformly in this analysis, may oversimplify the complex interplay between factors. Future research should explore {\bf data-driven weighting methods} to refine and optimize SCVI calculations, ensuring a more accurate representation of the dynamics influencing cyber vulnerabilities.

{\bf Fifth}, the cross-sectional nature of the datasets further limits the ability to draw causal interpretations and obscures how vulnerabilities evolve. Longitudinal studies are needed to capture dynamic changes in cyber vulnerabilities and provide deeper insights into temporal trends. Furthermore, sociocultural variations in trust cues and persuasion tactics restrict the generalizability of SCVI to diverse populations. Addressing these {\bf cultural specificities} in future research is essential to enhance the framework's applicability across global contexts.

{\bf Lastly}, the {\bf continuous evolution of cybercriminal tactics} necessitates regular updates to the SCVI framework to incorporate emerging scam vectors and adapt to the rapidly changing cyber landscape. 

These limitations highlight critical areas for refinement and underscore the need for future research to improve the SCVI framework's adaptability, scalability, and robustness as a comprehensive metric for assessing social cyber vulnerabilities.

\subsection{Future Work} \label{sec:future-work}

To address the identified limitations of the {\bf SCVI framework} and build on its strengths, several key future research directions are proposed. 

{\bf First}, the SCVI offers a holistic perspective on cyber vulnerability by integrating individual-level risk factors with contextual attack severity. Results from the iPoll and Reddit datasets illustrate that while many individuals maintain moderate or even low self-assessed susceptibility, they often encounter sophisticated or severe scams. This mismatch underscores {\bf the need for targeted interventions} addressing user vulnerability and the high-impact nature of many prevalent cyber threats. SCVI’s modular design, encompassing dimensions such as awareness, psychological traits, past experiences, and scam sophistication, facilitates more robust analyses than traditional metrics like CVSS and SVI, primarily focusing on software exploits or broad social indicators.

{\bf Second}, efforts should improve {\bf the reliability of input data} by reducing biases associated with self-reported data. Incorporating alternative data collection methods, such as behavioral tracking or third-party validation of reported incidents, can help mitigate recall bias and under-reporting. These methods would provide a more objective foundation for assessing vulnerabilities and enhancing SCVI's robustness.

{\bf Third}, {\bf expanding the geographic and demographic scope of SCVI} is essential for capturing a more diverse range of cultures, languages, and age groups. This would improve the framework's applicability and inclusivity. Leveraging diverse datasets across cultural and linguistic contexts could enhance SCVI's generalizability. Moreover, addressing underrepresented populations, such as older adults with limited Internet access and non-English speakers, would ensure a more comprehensive analysis of cyber vulnerabilities.

{\bf Fourth}, {\bf addressing sample size variability}, particularly in geographical analyses, is crucial for improving the reliability of SCVI estimates. Increasing sample sizes for underrepresented regions and states would reduce confidence intervals and enhance regional assessments. Stratified sampling methods or region-specific weighting schemes could also refine regional vulnerability analyses.

{\bf Fifth}, {\bf longitudinal studies} are needed to track individual and community-level changes in vulnerability over time. Such studies would provide valuable insights into how behaviors, awareness levels, and scam tactics evolve in response to emerging threats and mitigation strategies. Temporal analyses could facilitate the identification of trends and inform proactive intervention strategies.

{\bf Sixth, developing data-driven weighting methods} represents a promising direction for optimizing SCVI. Incorporating machine learning techniques, such as feature importance estimation or optimization algorithms, would enable dynamic adjustments of IVI and ASI weights based on real-time threat intelligence. These approaches would enhance SCVI’s predictive power and accuracy.

{\bf Seventh}, {\bf regular updates to SCVI} are critical to maintaining its relevance in an ever-evolving digital threat landscape. This includes incorporating emerging scam typologies such as AI-powered phishing and deepfake-related fraud. By adapting to new threats, SCVI can remain a versatile and contextually relevant framework for assessing cyber vulnerabilities.

{\bf Lastly, addressing cultural specificity} is essential for enhancing SCVI's global applicability. Sociocultural differences in trust cues, communication styles, and persuasion tactics must be explored to localize the framework. Comparative studies across regions provide deeper insights into cultural and behavioral factors, further refining SCVI.

These future research directions aim to enhance the SCVI framework's adaptability, robustness, and inclusivity. By expanding its geographic and demographic scope, integrating data-driven methods, and continuously updating the framework to reflect emerging threats, SCVI can become an indispensable tool for assessing and mitigating social cyber vulnerabilities in diverse and dynamic contexts.

\clearpage
\newpage
\section*{Ethics Considerations}

This research was conducted with careful attention to ethical principles, particularly regarding data privacy and the responsible use of publicly available datasets.

\begin{enumerate}
    \item \textbf{Use of the iPoll Dataset~\cite{ipoll-dataset2020}:} The iPoll dataset, which was utilized in this study, is publicly available and has already been anonymized to ensure that no personally identifiable information (PII) is present. This guarantees compliance with data privacy standards and ethical guidelines for the use of secondary data.

    \item \textbf{Reddit Scam Reports:} The Reddit scam reports used in this study were collected via web crawling from publicly accessible forums where individual users share experiences with scams.  The user identifiers in the crawled data were anonymized, and no attempts were made to de-anonymize or infer personal information about users. This aligns with ethical research practices and Reddit's terms of service regarding data usage. 

    \item \textbf{Compliance with Data Policies:}
    We ensured that our data collection and analysis process complied with the terms of use for both datasets. Additionally, care was taken to avoid data misuse that could harm individuals or groups.

    \item \textbf{Research Transparency and Reproducibility:}
    To promote transparency and reproducibility, we documented our data collection and preprocessing methods in detail while ensuring that any data shared as part of this research does not compromise the privacy or anonymity of individuals.
\end{enumerate}

By adhering to these measures, this study upholds the ethical standards expected in academic research and respects the privacy and rights of individuals whose information is included in the datasets.

\section*{Compliance with the Open Science Policy}

This study adheres to the Open Science Policy by ensuring transparency, accessibility, and reproducibility in all aspects of the research process. Below are the key measures undertaken:

\textbf{Data Transparency and Availability:} The datasets used in this study include the publicly accessible iPoll dataset and Reddit scam reports. The iPoll dataset is available from the Roper Center for Public Opinion Research and has been anonymized to remove personally identifiable information. The Reddit data, collected through web crawling, consists of user-generated content shared in public forums and was processed to ensure no privacy violations. Detailed descriptions of these datasets, including preprocessing steps, are provided in the appendices.

\textbf{Methodological Reproducibility:} To promote reproducibility, all analytical methods, modeling processes, and evaluation techniques employed in this study are detailed. The computational pipeline used to estimate the Social Cyber Vulnerability Index (SCVI) has been documented comprehensively, including sensitivity analysis and Monte Carlo simulations.

\textbf{Code Accessibility:} The codebase developed for calculating the SCVI, performing sensitivity analysis, and generating the results presented in this paper will be made available upon acceptance of this paper. It will be hosted on a public repository (e.g., GitHub) with proper documentation to facilitate reuse and further research.

\textbf{Ethical Compliance:} Ethical considerations, including data anonymization and adherence to data usage policies, were strictly followed. The details of these measures are outlined in the Ethics Considerations section.

By complying with these principles, this work aligns with the Open Science Policy, ensuring that the research is accessible, transparent, and reproducible, thereby fostering broader collaboration and advancing the field of cybersecurity research.

\bibliographystyle{IEEEtranSN}
\bibliography{ref}

\appendices

\begin{table*}[h!]
\centering
\caption{\sc Feature Extraction for Individual Vulnerability Index (IVI) from the iPoll Dataset}
\label{tab:ivi-calculation}
\footnotesize
\begin{tabular}{|P{8cm}|P{8cm}|}
\hline
 {\bf Related Questions}  & {\bf Response and Score Mapping}\\
\hline
\hline

\multicolumn{2}{|c|}{\bf Lack of Awareness $A_{1,k}$}     \\

\hline
Q7. Generally, how concerned, if at all, are you that you and/or a family member may fall victim to a scam? & Very concerned: 0, Somewhat concerned: 1, Not too concerned: 2, Not at all concerned: 3, DON’T KNOW/ SKIPPED ON WEB/REFUSED: ignore \\
\hline
Q8. How familiar were you with online romance scams; Q14: Grandparent scams? Q21: Government impostor scams; Q28 Census scams & Very: 0, Somewhat: 1, A little: 2, Not at all: 3, DON’T KNOW/ SKIPPED ON WEB/REFUSED: ignore \\
\hline
\multicolumn{2}{|c|}{\bf Lack Knowledge of Protect Measure $A_{2,k}$}     \\

\hline
Q37. Caller ID is a reliable way to know where a call is coming from? & True: 0; False: 5, Not sure: 3, SKIPPED ON WEB/REFUSED: ignore \\
\hline
Q38. When surfing the internet, it is always safe to interact with a website as long as the website has a locked box icon that indicates it is HTTPS secured. & Same as above \\
\hline
Q39. The IRS can call you about back taxes that you may owe without sending you a written notice first. & True: 5; False: 0, Not sure: 2, SKIPPED ON WEB/REFUSED: ignore \\
\hline
Q40. The Social Security Administration will contact you directly, either by phone or email, if there is a problem with your Social Security benefits. & Same as above \\
\hline
\multicolumn{2}{|c|}{\bf Frequency of behaviors increasing the risk of attack $B_{1,k}$}     \\
\hline
Q1. Not including time that you spend participating in online surveys, how often do you typically go online or access the Internet, including sending or receiving email? & Daily: 5, Several times a week: 4, several times a month: 3, Once a month: 2, Less than once a month: 1, Never: 0 \\
\hline
Q2. How often, if at all, do you use the Internet to do the following activities & Daily: 5, Several times a week: 4, several times a month: 3, Once a month: 2, Less than once a month: 1, Never: 0 \\
\hline
Q3. Have you ever done any of the following to meet potential dates or romantic partners at any point or time in your life? & Yes: 5, No: 0, Not sure: 3 \\
\hline

\multicolumn{2}{|c|}{\bf Trust level related to attack $P_{1,k}$}     \\
\hline
Q6. How well do the following statements describe you?(Overall, I expect more good things to happen to me than bad; I sympathize with others' feelings; I am a trusting person; Overall, I am pleased with my life.) & very well: 5, Somewhat: 3, Not at all: 0 \\
\hline
Q6. How well do the following statements describe you?(I find it difficult to get emotionally close to others; I worry a lot.) & very well: 0, Somewhat: 3, Not at all: 5 \\
\hline
\multicolumn{2}{|c|}{\bf Risk perception and impulsivity $P_{2,k}$}     \\
\hline
Q5. Have you ever developed a romantic relationship with someone that you have never met in person? & Yes: 5, No: 0, DON’T KNOW/ SKIPPED ON WEB/REFUSED: ignore \\
\hline
Q6. How well do the following statements describe you? (I tend to get involved in things that I later wish I could get out of; I tend to make up my mind quickly; I feel uneasy in social settings & Very well: 5, Somewhat: 3, Not at all: 0 \\
\hline
\multicolumn{2}{|c|}{\bf Past Experiences $E_{1,k}$}     \\
\hline
Q4. Thinking of the dates or romantic partners that you have met first online, have any of them ever done the following? (Lied about themselves, ask for money, etc.) & Yes: 5, No: 0, Not sure: 3 \\
\hline
Q9. To the best of your knowledge, have you ever been a target of a romance scam; Q15, Grandparent scam; Q22. Government impostor scams; Q29. Census scams; Q35. Identify theft&  Yes: 5, No: 0, Not sure: 3, SKIPPED ON WEB/REFUSED: ignore \\
\hline
Q12. To the best of your knowledge, has anyone you know ever been a target of a romance scam? Q19: grandparent scam. Q26. government impostor scams. Q33. Census scams & Yes: 0, No: 5, Not sure: 3, SKIPPED ON WEB/REFUSED: ignore \\
\hline
Q13. Did the person lose any money or suffer other financial losses due to the romance scam?  Q20: grandparent scam. Q27. government impostor scams. Q34. Census scams  & Same as above \\
\hline
Q36. Approximately, when did you experience identity theft?  & less then a year: 1, 1-2 year: 2, 3-5 year: 3,: 5-9 year: 4, more than 10 years: 5, don't know/skipped/refused: ignore \\
\hline
\multicolumn{2}{|c|}{\bf Responses to past incidents $E_{2,k}$}     \\
\hline
Q10. Have you ever lost money or suffered other financial losses due to a romance scam? Q17: grandparent scam. Q24. government impostor scams; Q31. Census scams  & Yes: 5, No: 0, DON’T KNOW/ SKIPPED ON WEB/REFUSED: ignore \\
\hline
Q11. Have you ever experienced any health problems or emotional distress due to a romance scam? Q18: grandparent scam. Q25. government impostor scams; Q32. Census scams  & Yes, health problems only: 3, Yes, emotional distress only: 3, both: 5, no: 0, DON’T KNOW/ SKIPPED ON WEB/REFUSED: ignore \\
\hline

\end{tabular}
\end{table*}

\begin{table*}[ht]
\centering
\caption{\sc Feature Extraction for the Attack Severity Index (ASI) from the iPoll Dataset.}
\label{tab:asi_calculation}
\footnotesize
\begin{tabular}{|p{8cm}|p{8cm}|}
\hline
\multicolumn{2}{|c|}{\bf Frequency Factor} \\
\hline
Q9. ``To the best of your knowledge, have you ever been a target of a romance scam?''    
& Yes: 5, No: 0, Not sure: 1, SKIPPED/REFUSED: ignore \\
\hline
Q12. ``To the best of your knowledge, has anyone you know ever been a target of a romance scam?''   
& Yes: 5, No: 0, Not sure: 1, SKIPPED/REFUSED: ignore \\
\hline
Q15. ``To the best of your knowledge, have you ever been a target of a grandparent scam?''  
& Yes: 5, No: 0, Not sure: 1, SKIPPED/REFUSED: ignore \\
\hline
Q19. ``To the best of your knowledge, has anyone you know ever been a target of a grandparent scam?''  
& Yes: 5, No: 0, Not sure: 1, SKIPPED/REFUSED: ignore \\
\hline
Q22. ``To the best of your knowledge, have you ever been a target of a government impostor scam?''   
& Yes: 5, No: 0, Not sure: 1, SKIPPED/REFUSED: ignore \\
\hline
Q26. ``To the best of your knowledge, has anyone you know ever been a target of a government impostor scam?''   
& Yes: 5, No: 0, Not sure: 1, SKIPPED/REFUSED: ignore \\
\hline
Q29. ``To the best of your knowledge, have you ever been a target of a Census scam?'' 
& Yes: 5, No: 0, Not sure: 1, SKIPPED/REFUSED: ignore \\
\hline
Q33. ``To the best of your knowledge, has anyone you know ever been a target of a Census scam?''  
& Yes: 5, No: 0, Not sure: 1, SKIPPED/REFUSED: ignore \\
\hline

\multicolumn{2}{|c|}{\bf Consequence Factor} \\
\hline
Q10. ``Have you ever lost money or suffered other financial losses due to a romance scam?''    
& Yes: 5, No: 0, Not sure: 1, SKIPPED/REFUSED: ignore \\
\hline
Q11. ``Have you ever experienced any health problems or emotional distress due to a romance scam?''    
& Yes, health only: 4; Yes, emotional distress only: 4; Yes, both: 5; No: 0, SKIPPED/REFUSED: ignore \\
\hline
Q13. ``Did the person lose any money or suffer other financial losses due to the romance scam?''    
& Yes: 5, No: 0, Not sure: 1, SKIPPED/REFUSED: ignore \\
\hline
Q17. ``Have you ever lost money or suffered other financial losses due to a grandparent scam?''    
& Yes: 5, No: 0, Not sure: 1, SKIPPED/REFUSED: ignore \\
\hline
Q18. ``Have you ever experienced any health problems or emotional distress due to a grandparent scam?''    
& Yes, health only: 4; Yes, emotional distress only: 4; Yes, both: 5; No: 0, SKIPPED/REFUSED: ignore \\
\hline
Q20. ``Did the person lose any money or suffer other financial losses due to the grandparent scam?''    
& Yes: 5, No: 0, Not sure: 1, SKIPPED/REFUSED: ignore \\
\hline
Q24. ``Have you ever lost money or suffered other financial losses due to a government impostor scam?''    
& Yes: 5, No: 0, Not sure: 1, SKIPPED/REFUSED: ignore \\
\hline
Q25. ``Have you ever experienced any health problems or emotional distress due to a government impostor scam?''    
& Yes, health only: 4; Yes, emotional distress only: 4; Yes, both: 5; No: 0, SKIPPED/REFUSED: ignore \\
\hline
Q27. ``Did the person lose any money or suffer other financial losses due to the government impostor scam?''    
& Yes: 5, No: 0, Not sure: 1, SKIPPED/REFUSED: ignore \\
\hline
Q31. ``Have you ever lost money or suffered other financial losses due to a Census scam?''    
& Yes: 5, No: 0, Not sure: 1, SKIPPED/REFUSED: ignore \\
\hline
Q32. ``Have you ever experienced any health problems or emotional distress due to a Census scam?''    
& Yes, health only: 4; Yes, emotional distress only: 4; Yes, both: 5; No: 0, SKIPPED/REFUSED: ignore \\
\hline
Q34. ``Did the person lose any money or suffer other financial losses due to the Census scam?''    
& Yes: 5, No: 0, Not sure: 1, SKIPPED/REFUSED: ignore \\
\hline

\multicolumn{2}{|c|}{\bf Sophistication Factor} \\
\hline
Q10, Q13, Q17, Q20, Q24, Q27, Q31, Q34 
& Yes: 5, No: 0, Not sure: 1, SKIPPED/REFUSED: ignore \\
\hline
\end{tabular}
\end{table*}

\begin{table*}[t]
\centering
\caption{\sc State-wise Summary of Mean IVI, ASI, SCVI, and Confidence Intervals}
\label{tab:state_summary}
\footnotesize
\begin{tabular}{@{}lcccccc@{}}
\toprule
\textbf{State} & \textbf{Sample Size} & \textbf{Mean IVI} & \textbf{Mean ASI} & \textbf{Mean SCVI} & \textbf{CI Lower} & \textbf{CI Upper} \\ \midrule
Alabama         & 21  & 2.2017 & 1.1519 & 1.7487 & 1.2129 & 2.2844 \\
Alaska          & 2   & 2.6621 & 2.5000 & 2.6511 & 0.4127 & 4.8894 \\
Arizona         & 52  & 2.4559 & 1.3683 & 1.9859 & 1.6792 & 2.2927 \\
Arkansas        & 14  & 2.2154 & 1.5321 & 1.8738 & 1.4236 & 2.3240 \\
California      & 197 & 2.3034 & 1.4641 & 1.9769 & 1.8126 & 2.1411 \\
Colorado        & 43  & 2.3126 & 1.2260 & 1.9192 & 1.5396 & 2.2989 \\
Connecticut     & 648 & 2.2852 & 1.0578 & 1.7606 & 1.6762 & 1.8451 \\
Delaware        & 8   & 2.2472 & 0.4125 & 1.3298 & 1.2127 & 1.4470 \\
District of Columbia & 8 & 1.9101 & 0.4538 & 1.1819 & 0.8204 & 1.5435 \\
Florida         & 105 & 2.3152 & 1.4808 & 2.0194 & 1.7806 & 2.2583 \\
Georgia         & 33  & 2.3542 & 1.5691 & 2.1189 & 1.6624 & 2.5753 \\
Hawaii          & 5   & 2.4058 & 1.1220 & 1.7639 & 1.0852 & 2.4426 \\
Idaho           & 13  & 2.1801 & 1.1246 & 1.8311 & 1.0335 & 2.6286 \\
Illinois        & 67  & 2.2632 & 1.3351 & 1.8928 & 1.6040 & 2.1815 \\
Indiana         & 38  & 2.3518 & 1.2792 & 1.8571 & 1.5881 & 2.1261 \\
Iowa            & 24  & 2.1667 & 1.0788 & 1.7283 & 1.2274 & 2.2291 \\
Kansas          & 17  & 2.4039 & 1.4394 & 1.9824 & 1.5044 & 2.4603 \\
Kentucky        & 12  & 2.1597 & 1.1825 & 1.6711 & 1.1725 & 2.1696 \\
Louisiana       & 16  & 2.1480 & 1.2581 & 1.7031 & 1.2765 & 2.1297 \\
Maine           & 10  & 2.3044 & 1.3910 & 1.9607 & 1.2731 & 2.6483 \\
Maryland        & 21  & 2.2142 & 1.1205 & 1.7728 & 1.2382 & 2.3074 \\
Massachusetts   & 39  & 2.3466 & 0.9431 & 1.7233 & 1.4070 & 2.0396 \\
Michigan        & 51  & 2.1960 & 1.3165 & 1.8337 & 1.5217 & 2.1457 \\
Minnesota       & 25  & 2.4252 & 1.4620 & 2.1542 & 1.6321 & 2.6763 \\
Mississippi     & 5   & 2.3783 & 2.0000 & 2.2781 & 0.9182 & 3.6381 \\
Missouri        & 41  & 2.3160 & 1.4295 & 1.9827 & 1.6268 & 2.3387 \\
Montana         & 10  & 2.1225 & 1.4240 & 1.8202 & 1.0690 & 2.5715 \\
Nebraska        & 26  & 2.2278 & 1.0808 & 1.6977 & 1.3492 & 2.0463 \\
Nevada          & 11  & 2.3484 & 2.3745 & 2.4792 & 1.7692 & 3.1891 \\
New Hampshire   & 3   & 2.1882 & 1.4300 & 1.8091 & 1.6056 & 2.0126 \\
New Jersey      & 31  & 2.1925 & 1.4452 & 1.9661 & 1.5002 & 2.4320 \\
New Mexico      & 14  & 2.2762 & 1.5393 & 2.0750 & 1.3655 & 2.7845 \\
New York        & 71  & 2.1837 & 1.0558 & 1.6327 & 1.4244 & 1.8410 \\
North Carolina  & 55  & 2.2854 & 1.4275 & 1.9378 & 1.6379 & 2.2377 \\
North Dakota    & 3   & 2.4277 & 1.5400 & 1.9838 & 0.9936 & 2.9741 \\
Ohio            & 57  & 2.3874 & 1.2839 & 1.9111 & 1.5969 & 2.2253 \\
Oklahoma        & 658 & 2.3456 & 1.1328 & 1.8324 & 1.7468 & 1.9179 \\
Oregon          & 20  & 2.2713 & 0.8440 & 1.6130 & 1.2306 & 1.9954 \\
Pennsylvania    & 666 & 2.2713 & 1.0643 & 1.7417 & 1.6594 & 1.8240 \\
Rhode Island    & 6   & 2.3661 & 2.3817 & 2.7535 & 1.2373 & 4.2697 \\
South Carolina  & 24  & 2.2538 & 1.2692 & 1.8270 & 1.3790 & 2.2750 \\
South Dakota    & 5   & 2.3018 & 0.4620 & 1.3819 & 1.2523 & 1.5115 \\
Tennessee       & 29  & 2.4633 & 1.2490 & 2.0675 & 1.5640 & 2.5710 \\
Texas           & 104 & 2.2039 & 1.1030 & 1.7012 & 1.5040 & 1.8984 \\
Utah            & 12  & 2.5477 & 1.5483 & 2.1577 & 1.4140 & 2.9014 \\
Vermont         & 465 & 2.3695 & 1.1945 & 1.8394 & 1.7480 & 1.9308 \\
Virginia        & 36  & 2.2987 & 0.9575 & 1.7331 & 1.3814 & 2.0848 \\
Washington      & 671 & 2.3105 & 1.2937 & 1.8845 & 1.7982 & 1.9708 \\
West Virginia   & 10  & 2.1353 & 0.5610 & 1.3481 & 1.0176 & 1.6787 \\
Wisconsin       & 48  & 2.2806 & 0.6758 & 1.5271 & 1.2690 & 1.7851 \\
Wyoming         & 5   & 2.5281 & 1.0560 & 1.7921 & 1.0546 & 2.5296 \\ \bottomrule
\end{tabular}
\end{table*}

\end{document}